\documentclass[10pt]{iopart}
\usepackage{subfigure}
\usepackage{graphicx}
\usepackage{color}
\usepackage[absolute,overlay]{textpos}
\usepackage{cancel}
\usepackage[mathscr]{euscript}
\usepackage{textcomp}
\usepackage{lmodern}
\usepackage[T1]{fontenc}
\usepackage[normalem]{ulem}
\graphicspath{ {./Figures/} }

\newcommand{\bv}[1]{\ensuremath{\mathbf{#1}}}

\newcommand{\unit}[1]{\ensuremath{\hat{#1}}}

\newcommand{\curl}[0]{\nabla \times}

\begin{document}

\title{Statistical analysis of \textit{m/n} = 2/1 locked and quasi-stationary modes with rotating precursors at DIII-D}
\author{R.~Sweeney\textsuperscript{1}, W.~Choi\textsuperscript{1}, R.J.~La~Haye$^2$, S.~Mao\textsuperscript{3,+}, 
K.E.J.~Olofsson\textsuperscript{1,*}, F.A.~Volpe\textsuperscript{1}, and the DIII-D~Team}

\address{\(^1\) Columbia University, New York, NY 10027}

\address{\(^2\) General Atomics, San Diego, CA 92121}

\address{\(^3\) University of Wisconsin, Madison, WI 53706}

\address{\(^+\) Present address: Stanford University, Stanford, CA 94305}

\address{\(^*\) Present address: General Atomics, San Diego, CA 92121}
\begin{abstract}

A database has been developed to study the evolution, the nonlinear
effects on equilibria, and the disruptivity of locked and
quasi-stationary modes with poloidal and toroidal mode numbers $m=2$
and $n=1$ at DIII-D. The analysis of 22,500 discharges shows that more
than 18\% of disruptions are due to locked or quasi-stationary modes
with rotating precursors (not including born locked modes). A parameter formulated by the plasma internal inductance
$l_i$ divided by the safety factor at 95\% of the poloidal flux, 
$q_{95}$, is found to exhibit predictive capability over
whether a locked mode will cause a disruption or not, and does so up to hundreds
of milliseconds before the disruption. Within 20 ms of the disruption, the shortest distance between the island
separatrix and the unperturbed last closed flux surface, referred to
as $d_{edge}$, performs comparably to $l_i/q_{95}$ in its ability to
discriminate disruptive locked modes. Out of all parameters considered, $d_{edge}$ also correlates best with the duration of the locked mode. 
Disruptivity following a $m/n=2/1$ locked mode as a function of the
normalized beta, $\beta_N$, is observed to peak at an intermediate
value, and decrease for high values. The decrease is
attributed to the correlation between $\beta_N$ and $q_{95}$ in the
DIII-D operational space. Within 50 ms of a locked mode disruption,
average behavior includes exponential growth of the $n=1$ perturbed
field, which might be due to the 2/1 locked
mode. Surprisingly, even assuming the aforementioned 2/1 growth,
disruptivity following a locked mode shows little dependence on island width up to 20 ms
before the disruption. Separately, greater
deceleration of the rotating precursor is observed when the wall torque
is large. At locking, modes are often observed to
align at a particular phase, which is likely related to a residual error field. Timescales
associated with the mode evolution are also studied and dictate
the response times necessary for disruption avoidance and mitigation. 
Observations of the evolution of $\beta_N$ during a locked
mode, the effects of poloidal beta on the saturated width, and the
reduction in Shafranov shift during locking are also presented.

\end{abstract}


\ioptwocol

\section{Introduction}

Rotating and non-rotating ("locked") neoclassical tearing modes (NTMs)
are known to degrade confinement and cause disruptions in tokamak
plasmas under certain plasma conditions, and thus represent a concern for ITER \cite{iterMHD}. 

NTMs can rotate at the local plasma rotation velocity, apart from a small
offset of the order of the electron or ion diamagnetic
velocity \cite{islandProp}. In present tokamaks with strong torque injection, this corresponds to rotation
frequencies of several kHz. Magnetic islands can also steadily rotate
at frequencies of the order of the inverse resistive-wall time (tens
of Hz, typically), if a stable torque balance can be established at
that frequency. In this case, they are called Quasi Stationary Modes
(QSMs) \cite{snipes}. 

There are also NTMs that do not rotate at all, called Locked Modes (LMs). 
Some of these are the result of an initially rotating
NTM ("rotating precursor") decelerating and locking to the 
residual error field. The deceleration might be due to
the magnetic braking experienced by the rotating island in its
interaction with the eddy currents that it induces in the resistive
wall \cite{Fitzpatrick}. Other modes are "born locked", i.e. they form without a rotating
precursor, as a result of resonant error field penetration. 

It will be important to understand the onset, growth, saturation, and
stabilization of all these categories of rotating and non-rotating
NTMs, in order to maintain good confinement and prevent disruptions in
ITER.  Here we present an extensive analysis of QSMs and LMs 
with rotating precursors, which we will sometimes refer to as "initially
rotating locked modes", or IRLMs. The analysis was carried over
approximately 22,500 DIII-D \cite{luxon} plasma discharges, and restricted
to poloidal/toroidal mode numbers \textit{m/n}=2/1, because these are
the mode numbers that are most detrimental to plasma confinement in DIII-D
and most other tokamaks \cite{butteryPPCF}. QSMs and LMs of different $m/n$ (for example
3/2, occasionally observed at DIII-D) and LMs not preceded by rotating
precursors are not considered here and will be the subject of a
separate work. 

Previous works have described classical \cite{Rutherford,white} and neoclassical \cite{chang_TFTR,butteryPPCF,lahaye2006} tearing modes (TMs), and the torques acting on TMs \cite{Fitzpatrick}. Effects of TMs on confinement are described theoretically \cite{chang}. Summaries of error-field-penetration locked modes in DIII-D \cite{scoville}, disruption phenomenology in JET \cite{disruptionsJET}, and disruption observations across many machines \cite{schuller} have been reported. 

Statistics regarding the role of plasma equilibrium parameters, and MHD stability limits on disruptions have been reported at JET \cite{deVries2009}, and one work includes a detailed accounting of the instabilities preceding the disruptions \cite{deVries2011}. A comprehensive study of disruptivity as a function of various equilibrium parameters was conducted at NSTX \cite{gerhardt} followed by a work on disruption prediction including $n=1$ locked mode parameters \cite{gerhardtDetect}. A study of various disruption types on JT-60U, including those caused by tearing modes, has been reported \cite{Yoshino1994}.  A brief study of disruptivity as a function of the safety factor and the normalized plasma beta on DIII-D has been reported \cite{FNSF}. 

Machine learning approaches to disruption prediction have shown rather high levels of success at ASDEX Upgrade \cite{cannas,pautasso2002}, JET \cite{ratta}, and JT-60U \cite{yoshino}, and a predictor tuned on JET was ported to ASDEX Upgrade \cite{windsor}. Similarly, neural networks have been used for disruption prediction at TEXT \cite{hernandez}, and ADITYA \cite{sengupta}, and for predicting ideal stability boundaries on DIII-D \cite{wroblewski}. Discriminant analysis for disruption prediction was tested at ASDEX Upgrade \cite{zhang}. 

A statistical work on MAST investigated the differences between disruptive and non-disruptive LMs as a function of normalized plasma beta and the safety factor \cite{butteryMAST}. Recently a locked mode thermal quench threshold has been proposed based on equilibrium parameters from studies on JET, ASDEX Upgrade, and COMPASS \cite{deVries2016}. The automatic detection of locked modes with rotating precursors makes the disruption statistics presented here unique. In addition, this work includes basic observations of locked modes and their effects on equilibria.

\begin{figure}[t]
        \includegraphics[scale=0.9]{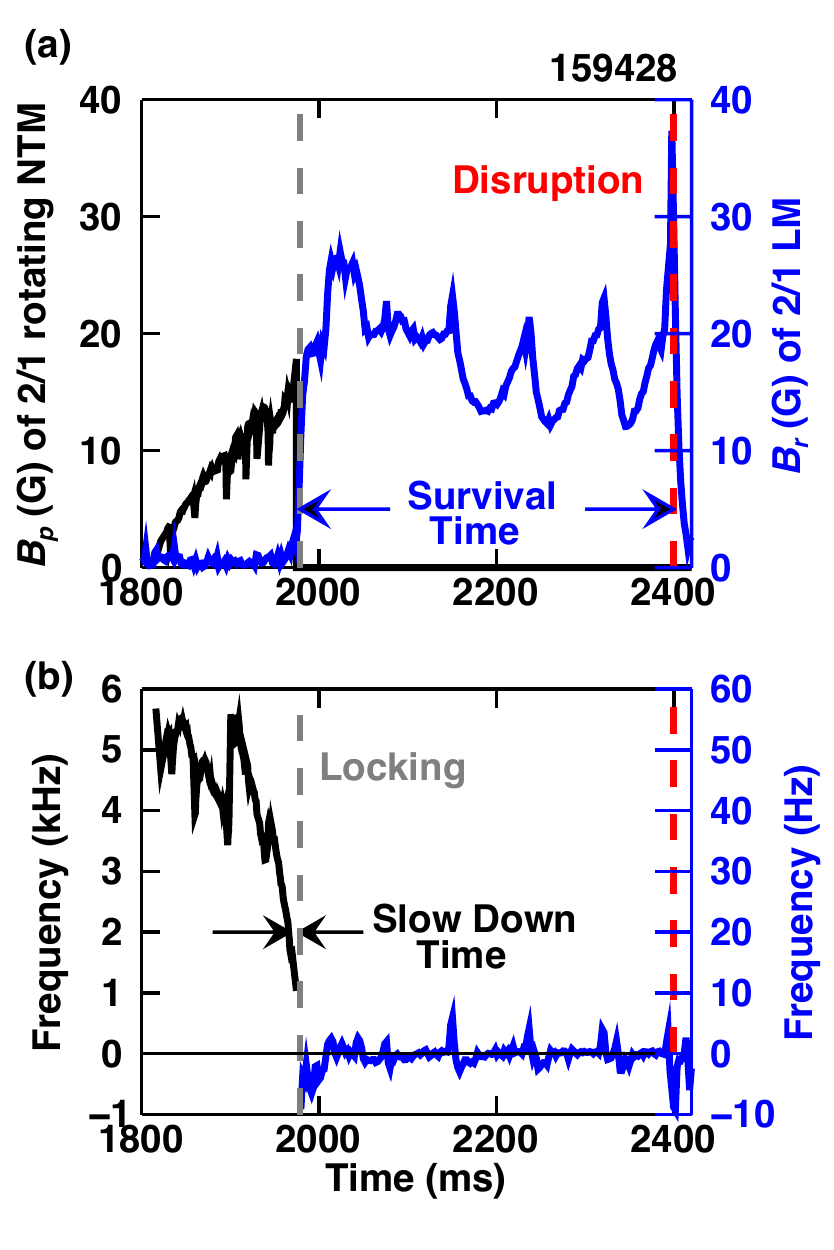} 
        \caption{Example of an initially rotating locked mode (IRLM). The black trace
        presents a fast rotating 2/1 NTM, as measured by the set of
        Mirnov probes and analyzed
        by \emph{eigspec}\cite{olofsson2014}. At the time of locking
        (1978.5 ms), the low frequency mode is detected by the ESLDs,
        shown in blue. 
        The slow-down time is the 
        time taken for a mode rotating at 2 kHz to slow and lock; 
        survival time is the duration of an IRLM that ends in a disruption.
        A factor of 0.5 has been multiplied to the
        Mirnov probes signal to account for the eddy currents in the
        wall during fast mode rotation. A factor of 2 has been multiplied to the
        ESLD signal to obtain the peak radial magnetic field from the
        measurement averaged over the large ESLD
        area.}  
\label{fig:eigspecExample}
\end{figure}

During the first discharges in this database, DIII-D was equipped with one poloidal and four toroidal arrays of Mirnov probes and saddle loops \cite{straitRSI}. During the time spanned by the database, additional sensors were added for increased 3D resolution \cite{kingRSI}. However, a limited set of six saddle loop sensors (external to the vessel) and three poloidal sensors (inside the vessel) are used for simplicity, and for consistency of the analysis across all shots, spanning the years 2005-2014. 

An example of the 2/1 IRLMs considered here is illustrated in
figure \ref{fig:eigspecExample}. The poloidal field amplitude of the
rotating precursor is detected by the toroidal array of Mirnov probes around
1800 ms. The mode simultaneously grows and slows down until it locks at 1978.5 ms. Due to the finite time binning 
 used in the Fourier analysis during the rotating phase, the rotating signal is lost at low frequency, and is instead
measured by a set of large saddle loops (ESLDs: external saddle loops differenced). The response of the saddle loops increases 
when $\omega < \tau_w^{-1}$, where
$\tau_w \sim3$ ms is the characteristic $n=1$ wall time for DIII-D. 
As shown in Fig.\ref{fig:eigspecExample}, 
it is not uncommon for the amplitude of an IRLM to oscillate due to minor 
disruptions, and to grow prior to disruption (this will be investigated in section \ref{sec:disruptiveGrowth}).

Three interesting results of this work are introduced
now. First, it will be shown that the $m/n=2/1$ island width cannot be used to distinguish disruptive from non-disruptive IRLMs 20 ms or more ahead of the disruption time. 
Similarly, the island width shows little correlation with the IRLM survival time.  

Second, the plasma internal inductance divided by the safety factor, $l_i/q_{95}$, distinguishes IRLMs that will disrupt from those that will not. The predictive capability of $l_i/q_{95}$ might be related to the energy available to drive nonlinear island growth. 

Finally, a spatial parameter which couples the
$q=2$ radius and the island width, referred to as $d_{edge}$ (see
section \ref{sec:survival} for definition), also distinguishes disruptive from non-disruptive IRLMs well within 20 ms of the disruption. It also
correlates best with the IRLM survival time. The
predictive capability of $d_{edge}$ is believed to be related to the
physics of the thermal quench.

The paper is organized as follows. Section \ref{sec:method} explains
the method of detection of disruptions, of rotating tearing modes, and of LMs. 
Section \ref{sec:survey} provides some general statistics of IRLM occurrences in DIII-D. 
Section \ref{sec:timescales} quantifies the timescales of interest 
before locking. Section \ref{sec:survival}
investigates the time available to intervene before an IRLM causes a disruption. 
Section \ref{sec:amplitude} discusses the width and phase behavior at locking, and the exponential growth of the $n=1$ field before the disruption. Section \ref{sec:beta} details the
interdependence between IRLMs and plasma $\beta$ ($\beta = \langle p \rangle / (B^2/2 \mu_0)$ where $\langle p \rangle$
is the average pressure and $B$ is the average total field strength). 
Section \ref{sec:rhoq2} decouples the  influence of $\rho_{q2}$, $q_{95}$, and $l_i$ on
IRLM disruptivity, and investigates the effectiveness of $l_i/q_{95}$, the island width, and $d_{edge}$ as disruption predictors. A discussion section follows which offers possible explanations of the physical relevance of $l_i/q_{95}$ and $d_{edge}$. Finally, two appendices are dedicated
to the mapping from radial magnetic field measurements to the perturbed island current, and from the perturbed current to an island width.

\section{Method}
\label{sec:method}

\subsection{Detection of disruptions}

To categorize disruptive and non-disruptive modes, a clear definition of disruption is needed.  The plasma current decay-time is used to differentiate disruptive and non-disruptive plasma discharges. The
decay-time \(t_D\) is defined as the shortest interval over which 60\% of the flat-top current is lost, divided by 0.6.
In cases where the monotonic decrease of \(I_p\) extends beyond 60\%, the entire duration of the current decrease is used, with proper normalization.
The disruption time is defined as the beginning of the current quench, which is usually preceded by a thermal quench, a few milliseconds prior.

The criterion $t_D <$40~ms used to identify DIII-D disruptions was formulated as follows. 

\begin{figure}
        \includegraphics[trim=0cm 0.5cm 0cm 0cm, clip=true, scale=0.95]{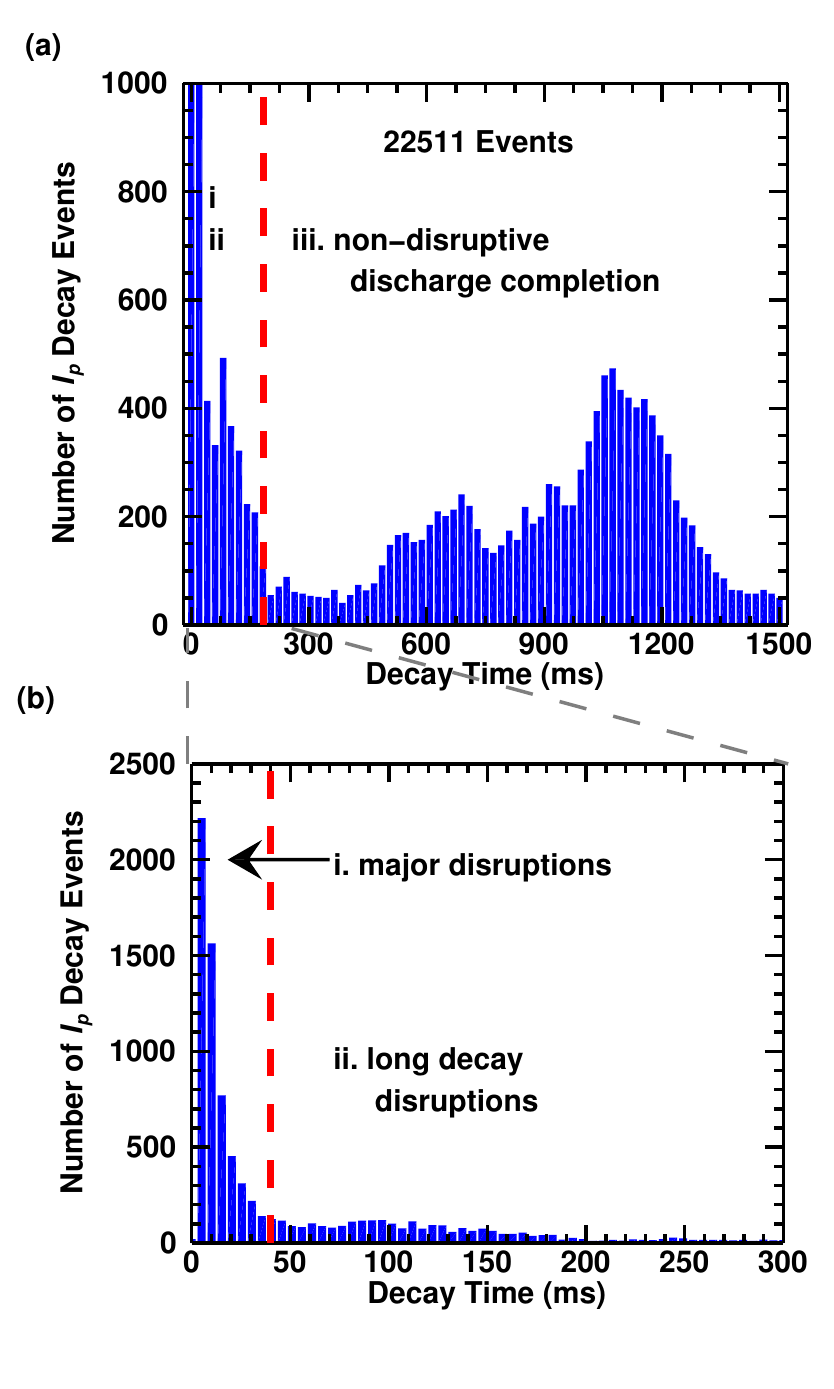} \caption{The
        distribution of plasma current decay time, roughly split into
        three populations. Panel (a) shows the non-disruptive
        discharges with decay times greater than 200 ms; panel (b)
        further distinguishes the remaining population into major
        disruptions ($<40$ ms, consisting of $>80\%$ flat-top disruptions), and disruptions with longer decay times,
        which are predominantly disruptions during ramp-down. Note that the vertical axis on panel (a) is
        interrupted to better show the features in the
        distribution.}  \label{fig:PlasmaCurrentDecayTime}
\end{figure}

A histogram of all decay-times is shown in
figure \ref{fig:PlasmaCurrentDecayTime}, and features of the distribution are used to define three populations.
The first group peaks near $t_D$=0 and extends up to $t_D$=40 ms. These are rapid losses of $I_p$ and confinement, 
quicker than typical energy and particle confinement times.  
Discharges in this group are categorized as major
disruptions, 
either occuring during the $I_p$ flat-top, or occurring during a partial controlled ramp-down of less than 40\% of the flat-top value. The sudden loss of current during the partial ramp-down cases must be fast enough to normalize to an equivalent 40 ms or less full current quench.
 It is worth noting that of the 5,783 disruptions detected,
666 occurred within the first second (in the ramp-up phase), none of which
were caused by a 2/1 IRLM.
At the opposite limit, group \textit{iii}, with 
\(t_D\) > 200 ms, contains mostly ($88\pm4\%$) non-disruptive discharges, 
in which the plasma current
decays at a steady rate for at least 80\% of the ramp-down. Population \textit{ii} has \(t_D\) in the range 40 ms < \(t_D\) < 200
ms, and mostly consists of shots that disrupted during the
current ramp-down with ``long decay'' times relative to population \textit{i} disruptions.

Note that while the stringent threshold of \(t_D\) < 40 ms will
prevent falsely categorizing non-disruptive discharges as disruptive,
it may also categorize some disruptions with
slightly longer decay times into group \textit{ii}. However, the thresholds are chosen to protect against false
positives better than false negatives; they are chosen to compromise
missing a number of disruptive shots in exchange for ensuring the
validity of all disruptive discharges. In the remainder of this work, we will focus on groups \textit{i} and \textit{iii} only.

In a manual investigation of 100 discharges in group \textit{i}, 85 disruptions occurred during the current flat-top, and 15 disruptions occurred during the current ramp-down phase.  Therefore, the majority of disruptions studied in this work (i.e. $85\pm4\%$) occur during a current flat-top.

In section \ref{sec:survey}, disruptivity will be studied over the entire database, including shots that did not contain IRLMs. We will refer to disruptivity in this context as \textit{global disruptivity} (i.e. the number of disrupted discharges divided by all discharges).

In all sections following section \ref{sec:survey}, disruptivity will be studied on shots that contained IRLMs only. In most cases, we will be interested in studying what differentiates a disruptive IRLM from a non-disruptive IRLM, and therefore we define \textit{IRLM disruptivity} as the number of disruptive IRLMs divided by the total number of IRLMs (where the total is the sum of disruptive and non-disruptive IRLMs). In one case, it will be useful to discuss \textit{IRLM shot disruptivity}, which is the number of disruptive IRLMs divided by the total number of \textit{discharges} with IRLMs (note that a non-disruptive discharge can have many non-disruptive IRLMs, making this distinction non-trivial).

\subsection{IRLM Disruptivity during current flat-tops}
\label{sec:irlmDisruptivity}
For all IRLM disruptivity studies, disruptions that occur during $I_p$ ramp-downs are limited to $15\pm4\%$ of the studied set. $I_p$ ramp-downs are characterized by major changes of the plasma equilibrium, and are expected to greatly impact the locked mode evolution. Namely, key parameters such as $q_{95}$, $l_i$, and $\rho_{q2}$ evolve during an $I_p$ ramp-down, complicating the interpretation of their effect on IRLM disruptivity. 
Moreover, flat-tops will be longer and longer in ITER and DEMO, and thus disruptivity during ramp-down will become less and less important. Eventually, in a steady-state powerplant, only flat-top disruptivity should matter. 

Out of 1,113 shots which disrupted due to an IRLM, 105 contained an additional IRLM distinct in time from the final disruptive one. As these additional IRLMs decayed or spun-up benignly, yet occurred in plasmas that ultimately disrupted, they are considered neither disruptive nor non-disruptive, and are excluded from the IRLM disruptivity studies. Similarly, a small number of discharges disrupt without an IRLM present, but contain an IRLM 100 ms before the disruption or earlier. In these cases, it is not clear whether the IRLM indirectly caused the disruption or not, and therefore these cases are also excluded.

\subsection{Detection of rotating modes of even $m$ and $n$=1}

Detection of a rotating mode is performed in two stages. For each
shot, the signals from a pair of toroidally displaced outboard midplane magnetic
probes are analyzed by the \emph{newspec} Fourier analysis
code \cite{straitRSI}, in search of \(n\)=1 activity. Genuine $n$=1
magnetohydrodynamic (MHD) activity is distinguished from $n$=1 noise  by
searching for both an $n$=1 amplitude sustained above a chosen
threshold, as well as requiring the corresponding $n$=1 frequency to
be "smooth". An adaptive threshold is used to accept both 
large amplitude, short duration modes, as well as small amplitude, long duration modes. 
Once detected, this  activity is analyzed by a simplified
version of the modal analysis \textit{eigspec} code, based on
stochastic subspace
identification \cite{olofsson2014}. Here \textit{eigspec} uses 9
outboard mid-plane and 1 inboard mid-plane Mirnov probes to isolate $n=1$ and determine
whether \(m\) is even or odd. Selecting modes of even $m$ and $n=1$
rejects $n=1$ false positives due to 1/1 sawtooth activity, and other odd $m$ and $n=1$ activity. Modes with $n$=1
and even $m$ are predominantly 2/1. Provided $q_{95}$ is
sufficiently high, they might in principle be 4/1 or 6/1 modes, but these modes
are rare. Manual analysis of 20 automatically
detected modes of even $m$ and $n$=1 found only 2/1
modes.

\subsection{Detection of $n=1$ locked modes}
\label{sec:detectLocked}

Locked modes are detected using difference pairs of the integrated external
saddle loops (ESLDs). A toroidal array of six external saddle loops is available. Differencing of loops positioned 180$^\circ$
apart toroidally eliminates all $n=$ even modes,  including the
equilibrium fields. A least squares approach is then used to fit the
$n=1$ and $n=3$ toroidal harmonics \cite{straitRSI}. 
This approach assumes that the contributions of odd $n\geq5$ are negligible.

Each pair-differenced signal is compensated for pickup of the non-axisymmetric coils, using a combination of analog and digital techniques. The accuracy of the coil compensations was assessed using
vacuum shots from 2011-2014. Residual coil pickup peaks at
$\sim3$ G, but a conservative threshold of 5 G is chosen for
identification of LMs  to avoid false positives. Small LMs that produce signals less than 5 G are
not considered in this work.

Analog integrators are known to add linear drifts to the saddle loop
signals.  In addition, $n=1$ asymmetries in
the plasma equilibrium can also produce background noise.

A simple yet robust algorithm was developed to subtract this
background.  The algorithm works on the principle that times exist
during which it is impossible for a LM to exist,  and the $n$=1
"locked mode signal" at those times must be zero.  These times include
the beginning and end of every shot, and  times at which $m/n=even/1$
modes are known, from Mirnov probe analysis, to rotate too rapidly to
be QSMs or LMs. As LMs cause a significant decrease in $\beta_N = \beta a B/I_p$ (as will be
discussed in  Section \ref{sec:beta}), the time when $\beta_N$ is
maximized is also highly unlikely to have a coincident LM,
and therefore, this time is also used. A piecewise linear function with nodes at each identified "2/1 LM 
free" time is fit to  each ESLD signal independently and subtracted to
produce a signal with minimal effects from integrator  drift and non-axisymmetric
equilibrium pickup.

Fifty shots automatically identified to have LMs with $n=1$ 
signals in excess of 5 G were manually analyzed. This analysis
confirmed that in most cases the automatic identification  was
accurate, with a percentage of false positives for LMs {\em
with} rotating precursors of $<4\%$.  The identification of locked
modes {\em without} rotating precursors (born locked modes), on the
other hand,  exhibited a percentage of false positives
$>30\%$. Greater accuracy is achieved for LMs with  rotating
precursors because the fast rotating precursor provides a LM free background
subtraction just prior to locking. In addition, locked modes 
with rotating precursors require two subsequent events: the appearance of a rotating
$m/n=even/1$ tearing mode followed by an $n=1$ locked mode. 
Due to the high percentage of false positives in the identification of born
LMs, they are not considered in this work, but will be the
topic of future work.

During the locked phase, no poloidal harmonic analysis is performed. It is
assumed that the confirmed $m/n=even/1$ rotating mode present
immediately before locking is likely a 2/1 mode, and upon
locking, the mode maintains its poloidal structure. In addition, it is
assumed that the locked $n=1$ signal measured by the ESLDs is
predominantly due to the 2/1 mode, such that this field
measurement can be used to infer properties of the mode. A set of 63 disruptive
IRLMs, occurring in plasmas with $|B_T|>2$ T, were investigated using
the 40 channel electron cyclotron emission (ECE) diagnostic \cite{ece}
to validate these assumptions. Only plasmas with $|B_T|>2$ T are
considered to ensure that ECE channels cover a plasma region extending from the core through
the last closed flux surface on the outboard midplane. Among these 63
IRLMs, 26 exhibited QSM characteristics, making full toroidal
rotations, allowing the island
O-point to be observed by the toroidally localized ECE diagnostic. In
all 26 cases, a flattening of the electron temperature profile is evident at the $q=2$ surface, and no surfaces with $q>2$ show
obvious profile flattening, suggesting that higher $m$ modes are not
present, or are too small to resolve with ECE channels separated by
$\sim1-2$ cm. 

As the gradient in the electron temperature
tends to approach zero near the core, confirming the presence or absence of a
$1/1$ mode is difficult. Despite the $1/1$ mode
existence being unknown, island widths derived from the radial field
measured by the ESLDs, where the $n=1$ signal is assumed to be a
result of the $2/1$ mode only, are calibrated to within $\pm2$ cm with the flattened $T_e$ profiles as measured by ECE (see Appendix B).  We conclude that in the majority of cases, the locked modes are 2/1 and the inferred island widths are reasonably accurate. The
minority of cases where this is not true are not expected to affect
the statistical averages presented in this work. 

The locked mode analysis includes a check for the existence of a $q=2$ surface, which is a necessary condition for the existence of a 2/1 IRLM. In 114 disruptive discharges, reconstructed equilibrium data are absent for 80\% or more of the locked phase, during which time the existence of a $q=2$ surface cannot be confirmed. The majority of these omitted discharges have locked phases lasting less than 20 ms, which is the time-resolution of equilibrium reconstructions, and therefore no equilibrium data exist after the mode locks (note that although 20 ms is a relatively short timescale, significant changes in the equilibrium are expected upon mode locking). All 114 discharges were manually analyzed, and approximately 40\% were identified as vertical displacement events or operator induced disruptions, such as discharges terminated by massive gas or pellet injection. About 15\% lack necessary data to manually identify the cause of the disruption. The remaining 53 disruptive discharges are considered valid 2/1 IRLM disruptions. These discharges are not included in the majority of the figures and discussion herein, but will be discussed in the disruption prediction section (section \ref{sec:prediction}) as they are expected to modestly decrease the performance of the predictors.

\subsection{Perturbed currents associated with the islands}

The mode amplitudes will sometimes be reported in terms of the total
perturbed current carried by the island $\delta I$.  $\delta I$ is a quantity that is local to the $q=2$ surface, and its calculation accounts for toroidicity. 
The wire filament model used
in \cite{shiraki} was shown to reproduce experimental  magnetics
signals well and was adapted for this calculation. An analytic version of
this model was developed  which simulates the island current perturbation
with helical wire filaments that trace out a torus of  circular
cross-section. The torus has the major and minor radii of the $q=2$
surface informed by experimental  EFIT MHD equilibrium reconstructions \cite{EFIT}, which use magnetics signals
and Motional Stark Effect measurements \cite{MSE} to constrain the reconstruction. An
analytic expression is found for $\delta I$ as a function of the
experimental measurement of $B_R$ from the ESLDs, and $R_0$  and $r_{q2}$ from EFIT 
reconstructions. The model and the
resulting analytic expression are detailed in Appendix A.

\section{Incidence of locking and global disruptivity on DIII-D}
\label{sec:survey}

\begin{figure}[h]
        \includegraphics[scale=0.65]{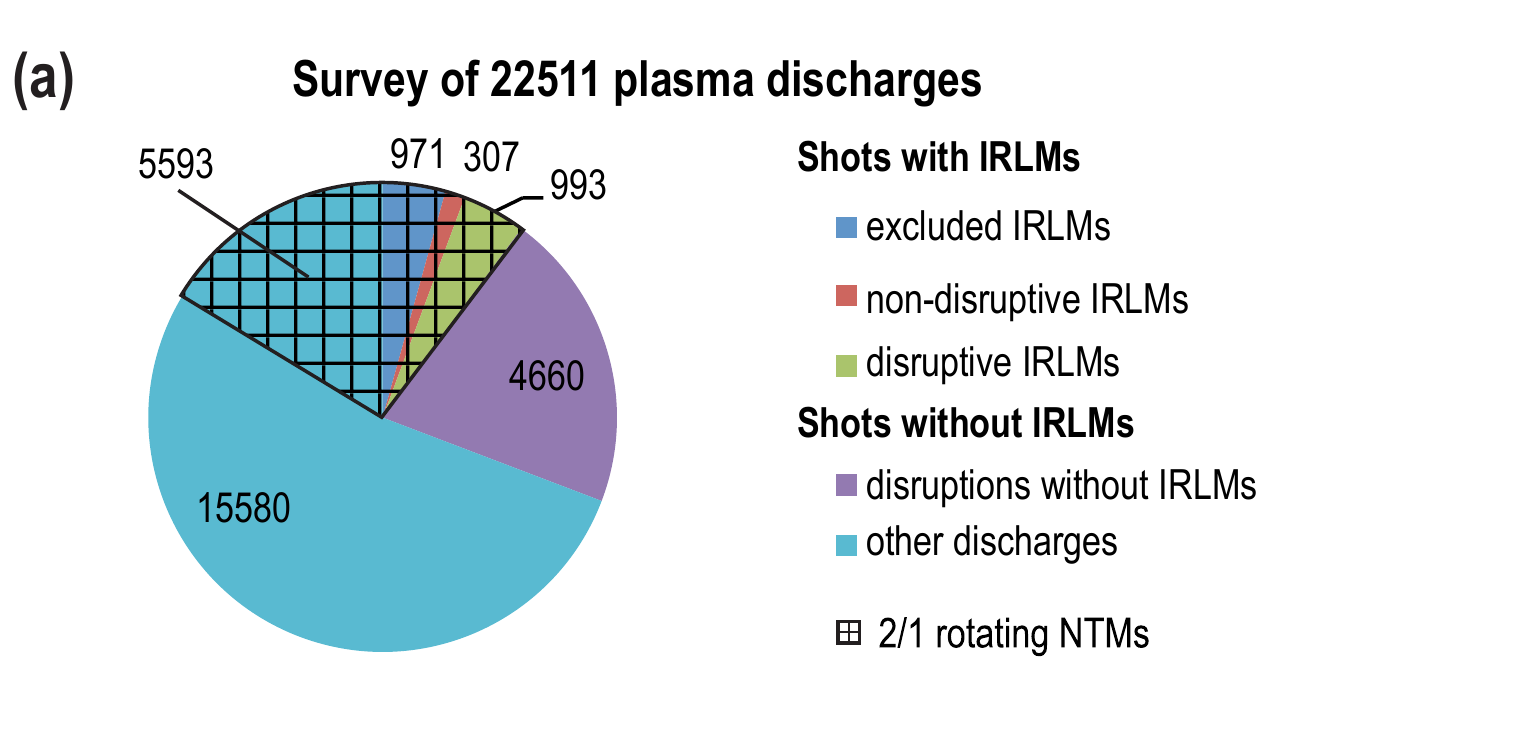} 
        \includegraphics[scale=0.65]{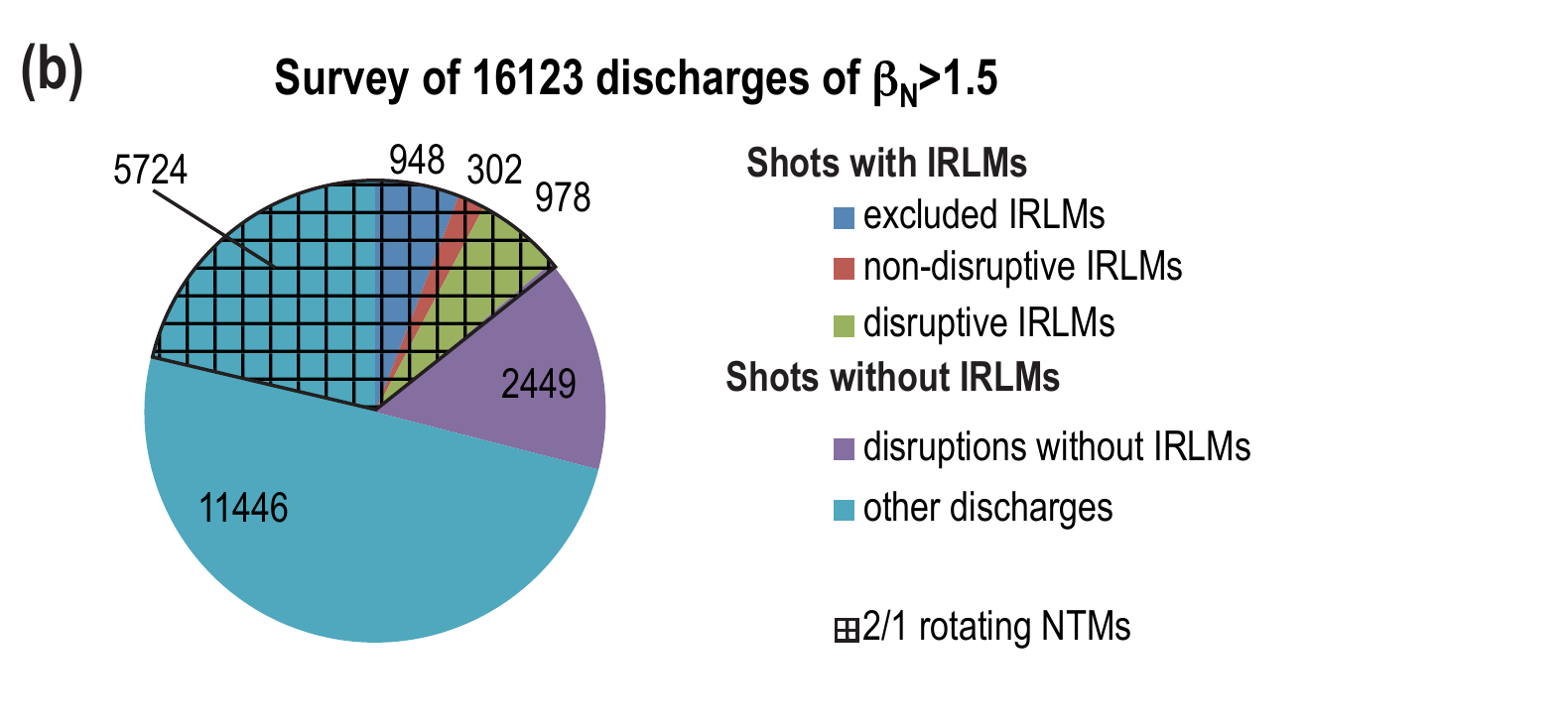} 
        \caption{(a)
          Color pie chart surveying all plasma discharges, showing the
          fraction of discharges with disrupting and non-disrupting
          initially rotating locked modes (IRLMs), as well as
          disruptions without IRLMs. Overplotted as a hatched region are the
          discharges with rotating 2/1 NTMs. (b) Same pie chart as (a), but
          for discharges with peak \(\beta_N >
          1.5\). Note that there is an overlap of 23 shots 
	between the hatched rotating NTM and the purple disruption regions.} 
\label{fig:DisruptionPieChart}
\end{figure}

To motivate the importance of study of these \(m/n\) = 2/1 modes,
figure \ref{fig:DisruptionPieChart} shows how often initially rotating
2/1 locked modes (IRLMs) occur in DIII-D plasmas. When considering all
plasma discharges, 25\% contain a 2/1 rotating NTM, 41\%
of which lock. Shots with IRLMs end in a major disruption 76\% of the time (using only the red and green portions of
figure \ref{fig:DisruptionPieChart}a, and excluding the blue portion). Approximately 18\%
of all disruptions are a result of an IRLM, in good agreement with the
$\sim16.5$\% reported on JET \cite{deVries2011}, and this statistic
rises to 28\% for shots with peak \(\beta_N\) > 1.5 (figure \ref{fig:DisruptionPieChart}b). The correlation between high \(\beta_N\) and rate of occurrence of IRLMs will be
detailed in section \ref{sec:betaDisruptivity}.

The blue slices show the number of IRLMs excluded
from the disruptivity studies, 
which consist of IRLMs in type \textit{ii} disruptions (long decay disruptions), IRLMs that terminate during a non-disruptive current ramp-down,
or IRLMs that cease to exist prior to a major disruption.
The ``other discharges'' do not contain IRLMs, and include long decay disruptions and non-disruptive discharges.

While the vast majority of rotating NTMs lock before causing a disruption, there were approximately 23 instances of the rotating 2/1 mode growing large enough to disrupt before locking. 

\section{Timescales of locking}
\label{sec:timescales}
In this Section we present two timescales indicative of the time 
available for intervention before locking. These timescales are useful for disruption avoidance and mitigation techniques.

\begin{figure}[t]
        \includegraphics[trim=0.2cm 0cm 0cm 0cm, clip=true, scale=0.48]{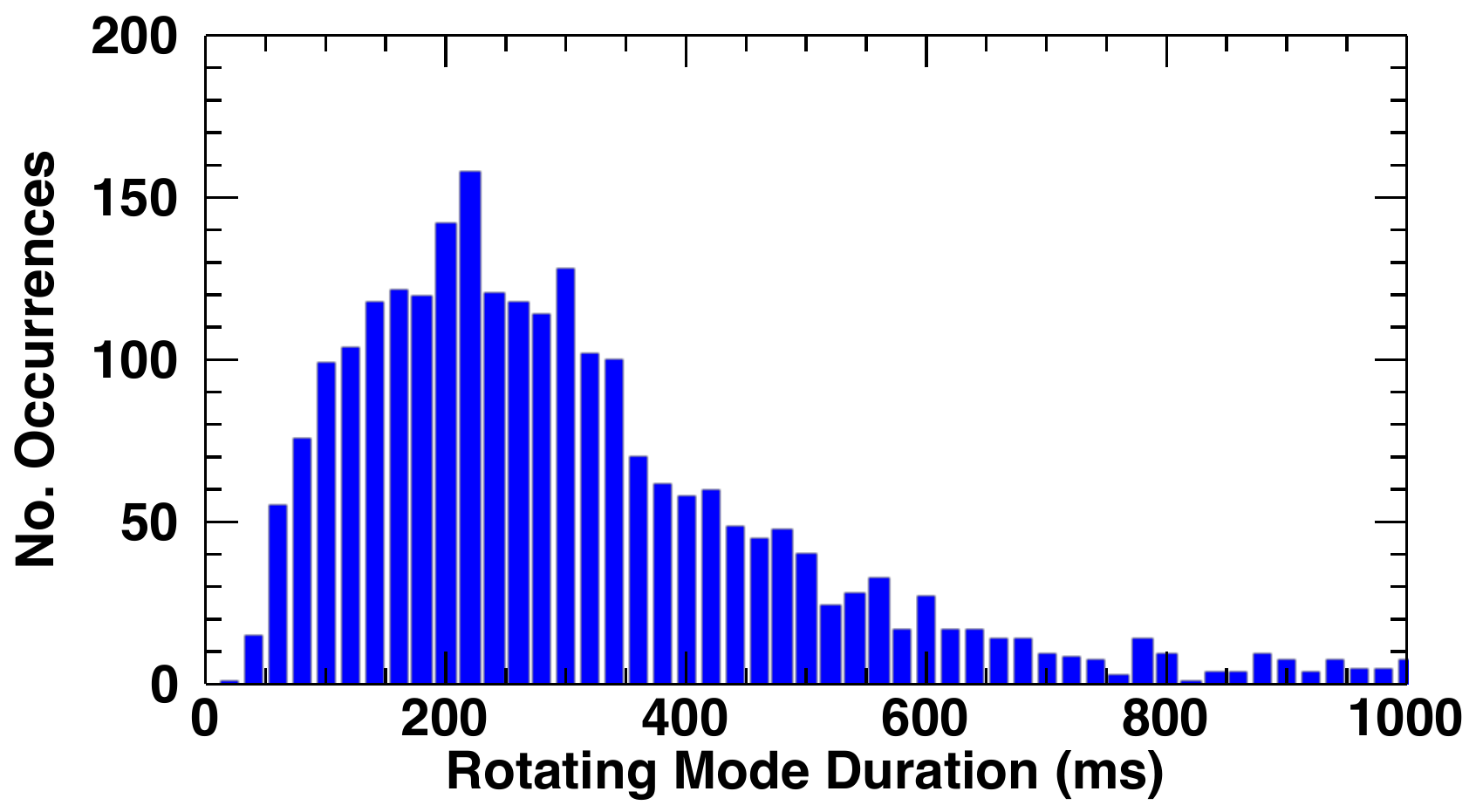} 
\caption{Histogram distribution of duration of rotating precursors (from 
mode onset to locking).}  
\label{fig:RotModeDuration}
\end{figure}

Figure \ref{fig:RotModeDuration} shows the duration of all
rotating \(m\)=even, \(n\)=1 modes that locked. 
A broad peak exists between 50 and 400 ms. The rotating duration can
depend on several different factors, such as the plasma rotation frequency, applied Neutral Beam Injection (NBI) torque, the island moment of inertia,
and viscous torques. The spread of values gives an indication of the time available to prevent locking, if an intervention is triggered upon rotating mode detection.

\begin{figure}[t]
        \includegraphics[trim=0.1cm 0cm 0cm 0cm, clip=true, scale=0.46]{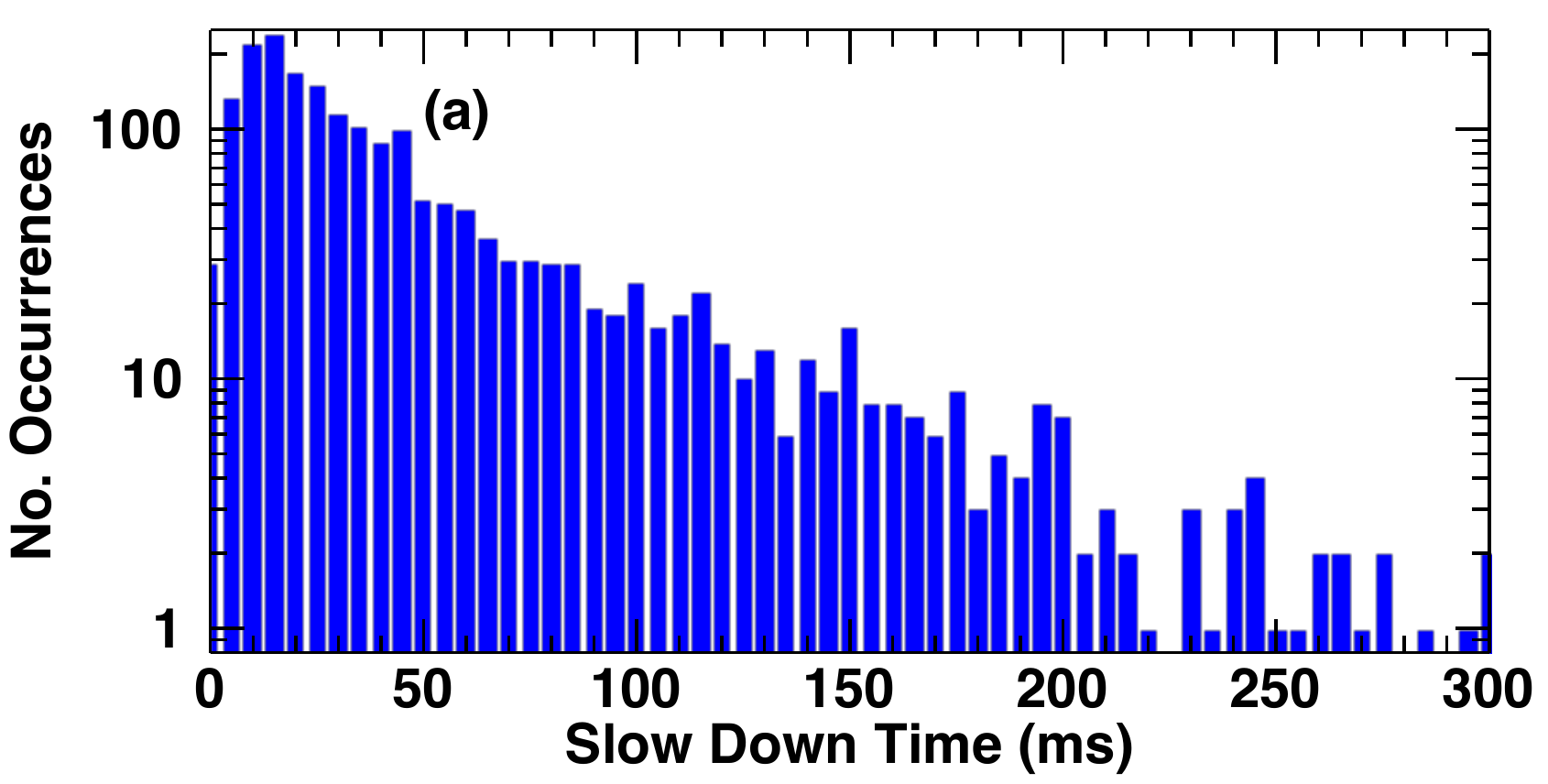} 
        \includegraphics[trim=0.1cm 0cm 0cm 0cm, clip=true, scale=0.42]{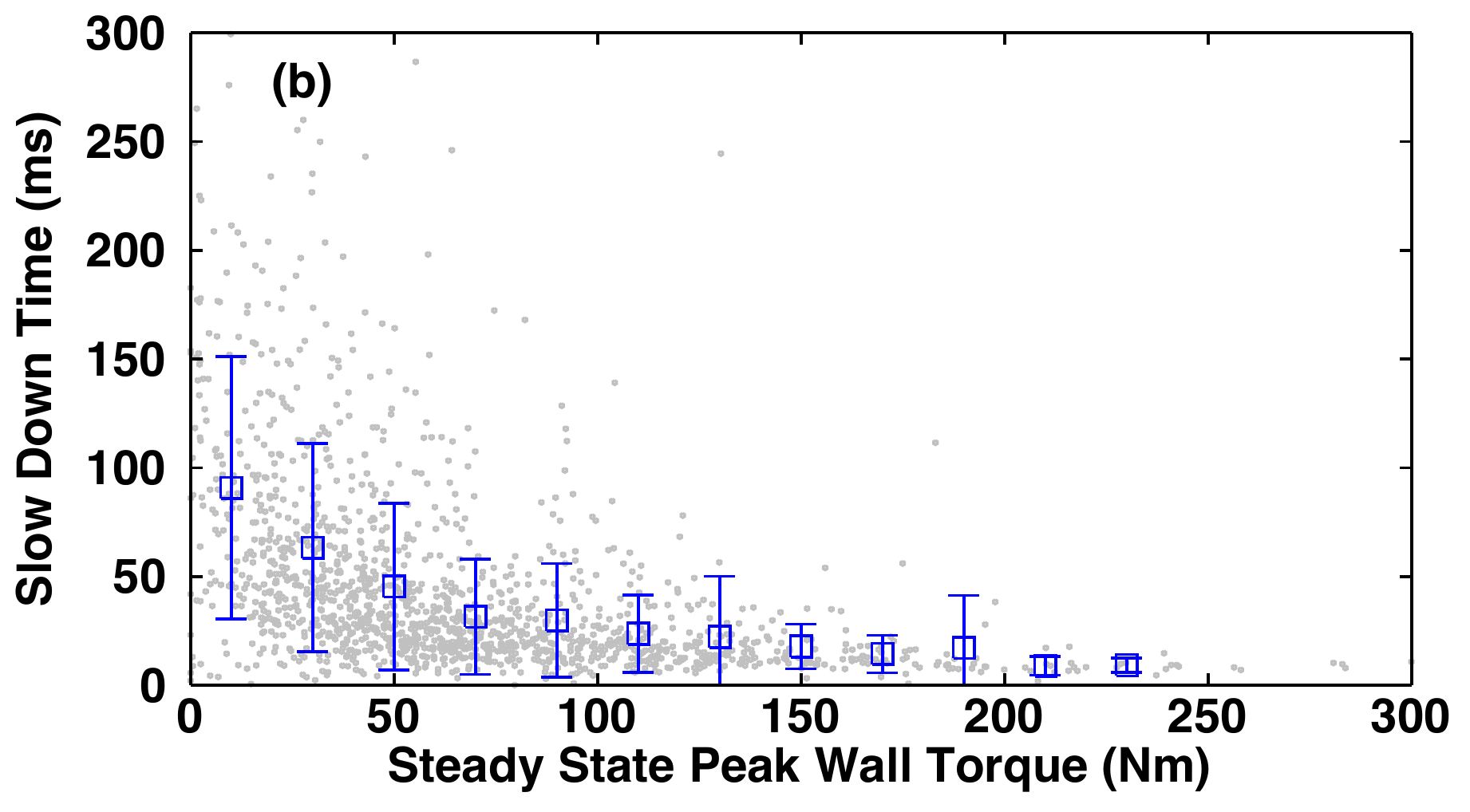} 
        \caption{(a)
        The time taken for a rotating \(m/n=2/1\) mode to slow from 2
        kHz to locked, as measured by \emph{eigspec} and ESLDs
        respectively. (b) A correlation is observed between the
        measured slow-down time and electromagnetic torque between the
        mode and the wall. 
	The torques are calculated by equation \ref{eq:wallTorque},
	where the perturbed magnetic field is taken when the mode is rotating at 2 kHz, and 
	\(\omega\tau_w\) is set to 1, representing the maximum of the frequency dependent steady-state wall torque.
	The points and error bars are the
        mean and standard deviations of each bin respectively. Note that
        about 5\% of the events lie beyond 300 ms, and are not plotted
        (in either panel). } 
\label{fig:SlowDownTime}
\end{figure}

The time taken for a mode rotating at 2 kHz to slow down and lock is referred to here as the slow-down time. The
threshold of 2 kHz was chosen empirically, as modes that decelerate to this 
frequency are often observed to lock. It is probable that at this frequency, the decelerating wall torque is stronger than the accelerating viscous torque, and causes the mode to lock. 
Figure \ref{fig:SlowDownTime}a shows 66\% of of slow-down times between 5 and 45 ms, with the peak of the distribution at $17\pm10$~ms; this is an
indication of the time available to prevent locking if measures are taken when the mode reaches 2 kHz.

Figure \ref{fig:SlowDownTime}b shows that modes which experience a larger wall torque generally slow down quicker than those with smaller wall torques. At low electromagnetic torque, the spread of slow-down times in figure \ref{fig:SlowDownTime}b suggests that other effects, such as the NBI torque, also become important. 

The toroidal electromagnetic torque between the rotating mode and the wall $T_{\phi,w}$ is expressed as follows \cite{Fitzpatrick},

\begin{equation}
T_{\phi, w} =  \frac{R_0 (2 \pi  r_s B_{rs} )^2}{\mu_0 n/m } 
\frac{ (\omega \tau_w)(r_{s+}/r_w)^{2m} }{1 + (\omega \tau_w)^2 [1 - (r_{s+}/r_w)^{2m}]^2 } 
\label{eq:wallTorque}
\end{equation}

\noindent where $m$ and $n$ are the poloidal and toroidal harmonics, $r_s$ is the minor radius of the $q=2$ surface, 
$r_{s+} = r_s + w/2$ with $w$ being the island width, $B_{rs}$ is the perturbed radial field at the $q=2$ surface,
$r_w$ is the minor radius of the resistive wall, and $\omega$ is the rotation frequency of the NTM co-rotating with the plasma. This form of 
the electromagnetic torque comes from a cylindrical approximation. 
The maximum of this torque occurs at the rotation frequency where $\omega \tau_w=1$, and is the quantity plotted on the horizontal axis of figure \ref{fig:SlowDownTime}b.

\section{Time between locking and disruption}
\label{sec:survival}

\begin{figure}[t]
        \includegraphics[trim=0cm 0.2cm 0cm 0cm, clip=true, scale=0.46]{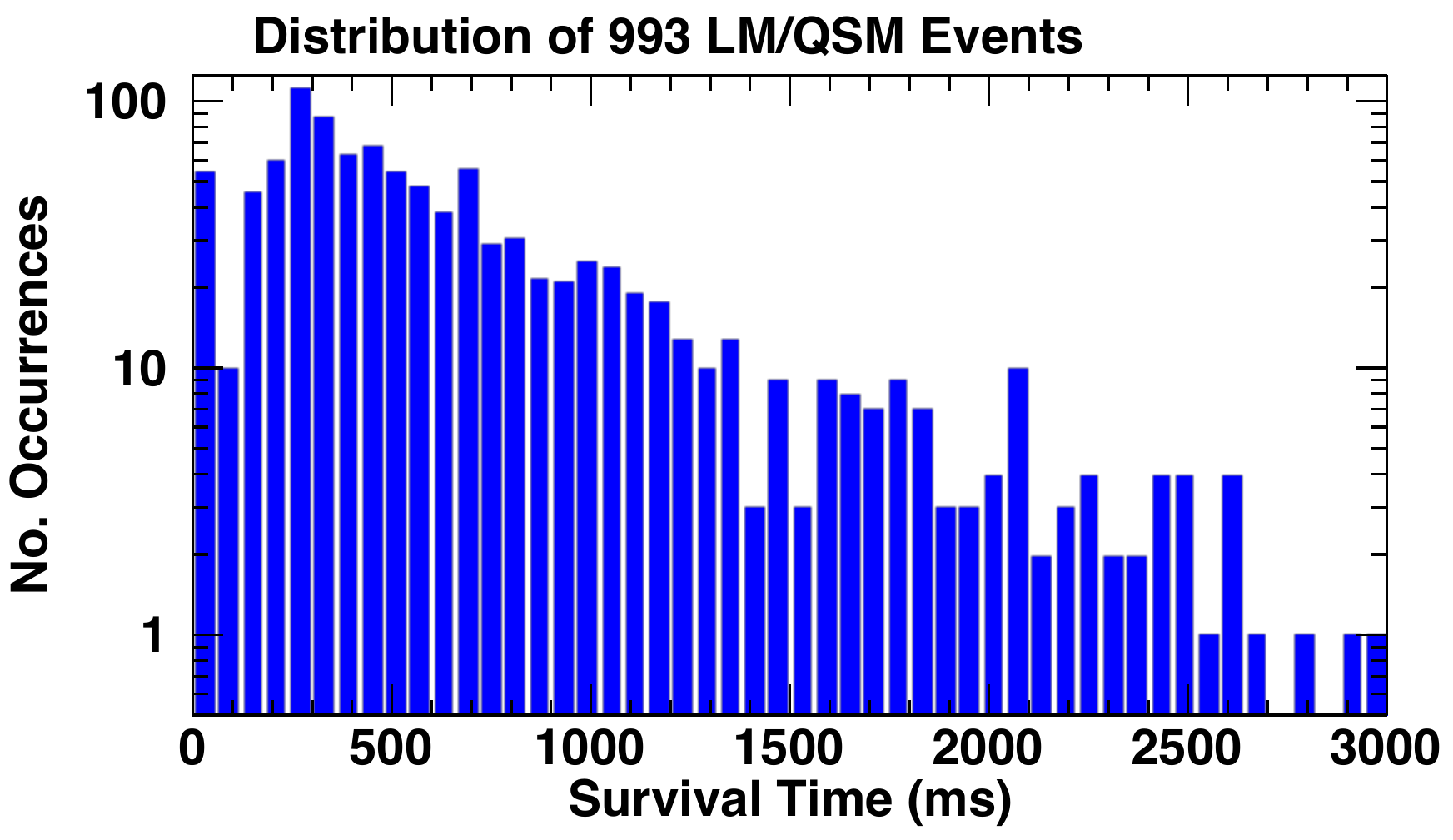} 
        \caption{
        A histogram of the survival time, defined as the duration of a
        locked mode that ended in a disruption. Less
        than 2\% of events survive for more 3000
        ms.}  \label{fig:SurvivalTime}
\end{figure}

The survival time is defined as the interval between locking and
disruption (note this is only defined for disruptive IRLMs). Figure \ref{fig:SurvivalTime} is a histogram of the
survival times of all disruptive IRLMs. Two groups can be observed, 
peaking at less than 60 ms and at 270 ms. 
The first group consists of 55 large rotating modes that lock and disrupt almost
immediately, as opposed to the latter spread of LMs that
reach a meta-stable state before disrupting. 
These short-lived modes, though dangerous and undesirable, 
could not be studied with the automated analysis as necessary equilibrium data
do not exist. These discharges are the same omitted discharges mentioned previously in the end of section \ref{sec:detectLocked}. Recall that these discharges are not included in the majority of the figures and discussion herein, but will be discussed in the disruption prediction section (section \ref{sec:prediction}) as they are expected to modestly decrease the performance of the predictors.

While 75\% of the population (excluding the 55 transient modes) survive between 150 to 1010 ms, the most frequent survival time is 270 \(\pm\) 60 ms, an indication of the time available to avoid disruption when a mode locks.

Gaining predictive capability over how long a disruptive locked mode is expected 
to survive might guide the best course of 
action to take, e.g. whether to stabilize the mode, or directly deploy disruption mitigation techniques. 

\begin{figure*}
        \includegraphics[trim=0.1cm 0cm 0cm 0cm, clip=true, scale=1.5]{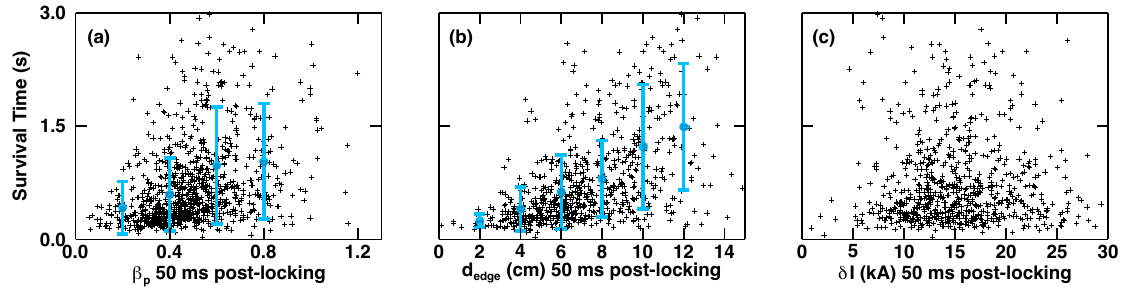} 
        \caption{Survival
        time shows some dependence on (a) $\beta_{p}$ and (b)
        $d_{edge}$.  (c) No
        correlation is found with $\delta I$ or similarly with the
        island width $w$ (see table \ref{tab:tsCorr}).
        } \label{fig:survTimeDependences}
\end{figure*}

Figure \ref{fig:survTimeDependences} shows the survival time plotted
against the poloidal beta $\beta_{p}$, the distance $d_{edge}$, 
and the perturbed island current $\delta I$. $d_{edge}$ is a quantity that measures the shortest
distance between the island separatrix and the unperturbed plasma separatrix: 
$d_{edge} \equiv a - (r_{q2} + w/2)$ where $a$ is the minor radius of
the unperturbed plasma separatrix, $r_{q2}$ is the minor radius of the $q=2$
surface, and $w$ is the island width.

\begin{table}[h]
	\centering
    \begin{tabular}{c | c}
    \textbf{Parameter} & \textbf{Correlation with $t_s$} \\
    \hline
    $d_{edge}$ & 0.47 \\
    $\rho_{q2}$ & -0.42 \\
    $l_i/q_{95}$ & -0.39 \\
    $q_{95}$ & 0.36 \\
    $\beta_p$ & 0.34 \\
    $dq/dr(r_{q2})$ & -0.15 \\
    $l_i$ & -0.11 \\
    $w$ & 0.10 \\
    $\delta I$ & -0.01 \\
    \end{tabular}
    \caption{Correlations of various parameters with the IRLM survival time $t_s$. The parameters are ordered in the table by the absolute value of their correlation coefficient. 
    Negative correlation means that a linear relationship with a negative slope exists between the parameters. } 
    \label{tab:tsCorr}
\end{table}

In figure \ref{fig:survTimeDependences}, the  survival time shows some
correlation  with $\beta_{p}$ and $d_{edge}$, but not with $\delta
I$. The correlation coefficients for these, and other
parameters are listed in table \ref{tab:tsCorr}.  The lack of correlation between
survival time and $\delta I$ is consistent with the lack of
correlation with the  island width $w$. This suggests that large
islands will not necessarily disrupt quickly, but islands which extend
near to the unperturbed separatrix (i.e. islands with small
$d_{edge}$) tend to disrupt quickly. The correlation of
$d_{edge}$ with survival time suggests that it is
pertinent to the physics of the thermal quench. Other works have found that parameters similar to $d_{edge}$ appear to cause the onset of the thermal quench \cite{henderCompass,Izzo,Sykes} (see section \ref{sec:discussion} for details). 

The best
 correlations in the table are considered moderate (i.e. moderate
 correlations are in the range $r_c=[0.4, 0.6]$, where $r_c$ is the
 correlation coefficient). The correlations in table \ref{tab:tsCorr} do not
 provide significant predictive capability. 
A macroscopic timescale like the survival time likely depends on many variables, and on the nonlinear evolution of the plasma under the influence of the locked mode. 
Note that $15\pm4\%$ of the disruptive IRLMs in this survival time study terminate during a plasma current ramp-down, and might have survived longer, had they not been interrupted.

\section{Mode amplitude and phase evolution}
\label{sec:amplitude}

\subsection{Distributions of IRLM toroidal phase at locking}
\label{sec:locking}

\begin{figure}[h]
	\centering
        \includegraphics[scale=0.6]{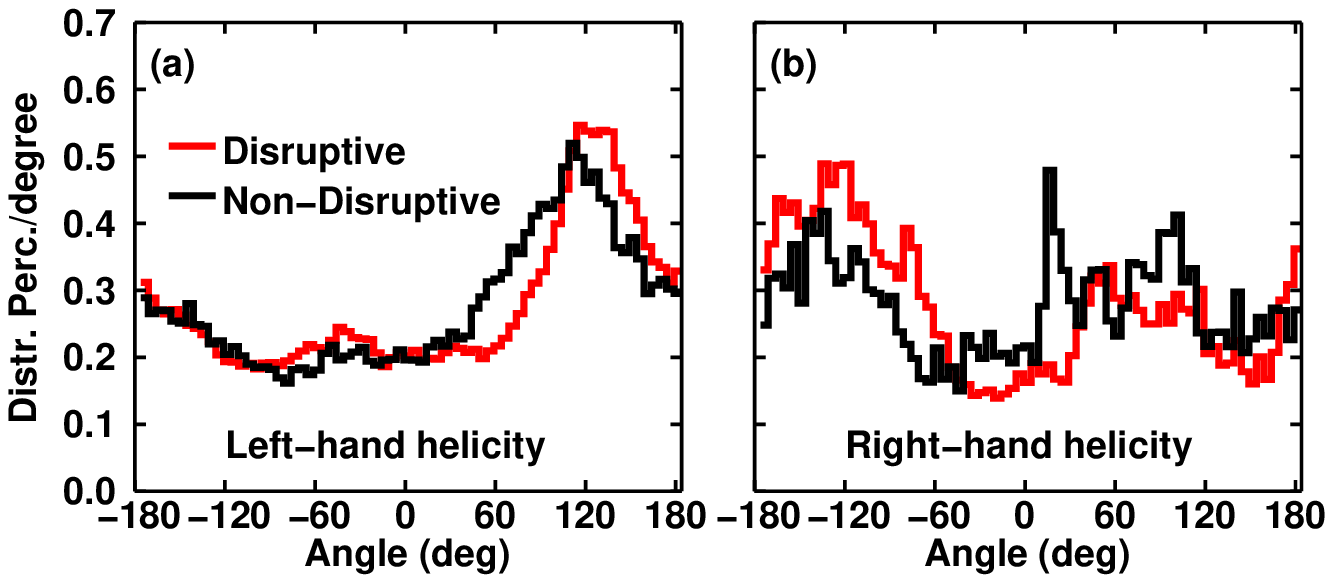}
        \includegraphics[scale=0.37]{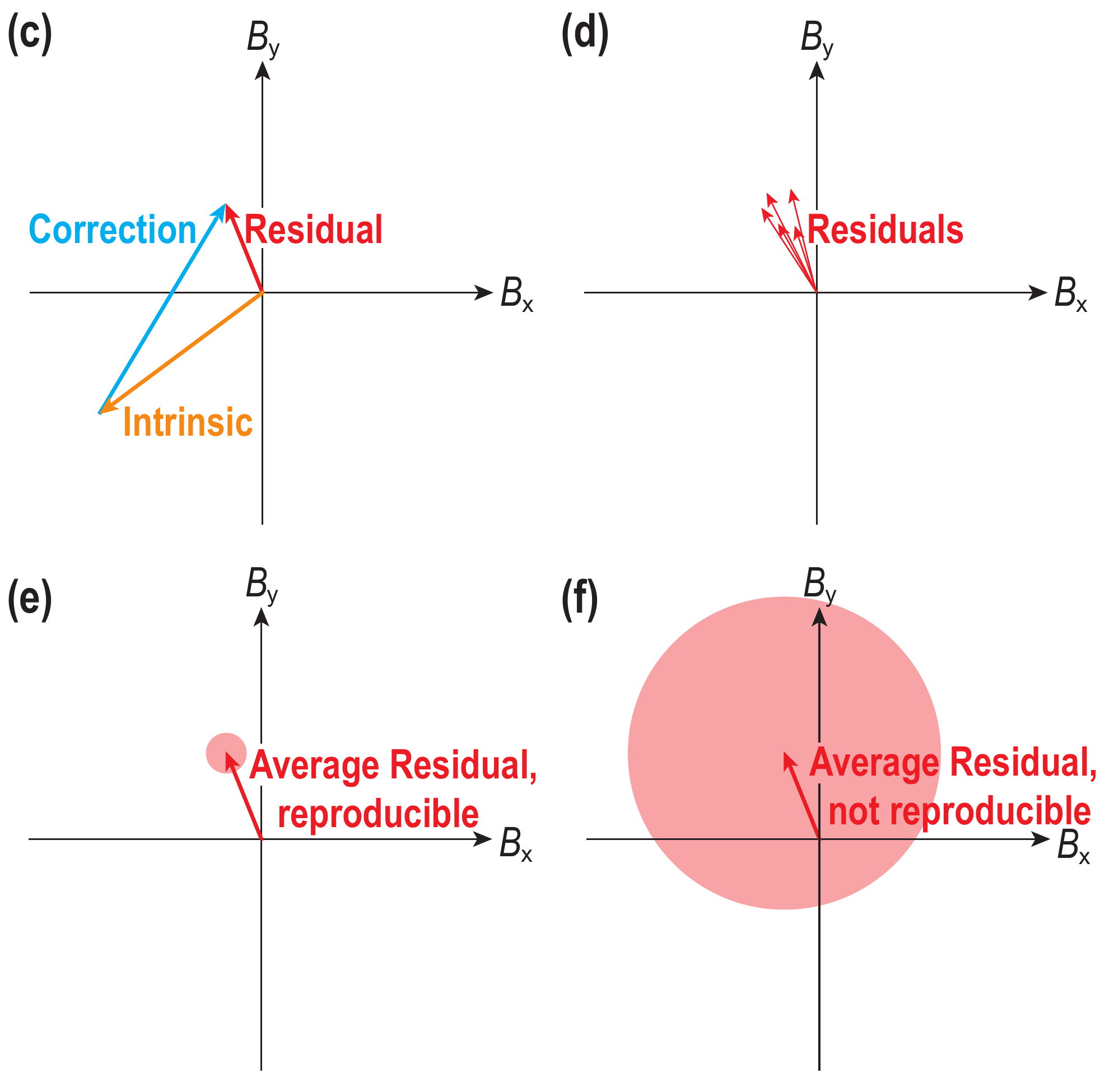}          
         \caption{
	The normalized
        phase distribution of all locked modes. Each mode contributes a total of $100/N$ across all
        bins ($N$ is number of disruptive or non-disruptive IRLMs).
	A binsize of 5 degrees was chosen as a compromise between 
        being large enough to have sufficient statistics in one bin and fine enough to show features of interest.
        The angle is where the radial field is largest and outward on the outboard midplane. 	
        (a) Left-hand helicity (i.e. normal $I_p$ and $B_T$, or both reversed) with 980 disruptive and 1029 non-disruptive IRLMs. 
	(b) Right-hand helicity plasma discharges (i.e. either $I_p$ or $B_T$ reversed). Only 130 disruptive and 204 non-disruptive IRLMs here.
	(c-f) Illustrations to explain distributions. 
        (c) The residual EF is the difference between the intrinsic EF and the applied correction. (d) Due to changes in the intrinsic and/or the correction, the residual EF can change. (e) An average residual can be defined by averaging over several shots and times. A small standard deviation in its amplitude and phase (illustrated by the small circle) are indicative of high reproducibility of the residual EF. In that case, a narrow distribution is expected in Fig.a-c. (f) In the opposite limit, the residual EF phasor can point to any quadrant, and a broad, flat distribution is expected        
        }
  \label{fig:PhaseDistribution}
\end{figure}

As a rotating $n=1$ mode is slowing down and about to lock, it tends to align
with existing \(n\) = 1 fields. In most cases, this will be the
residual error field, defined as the vector sum of the intrinsic error
field and the applied \(n\) = 1 error field
correction. Figure \ref{fig:PhaseDistribution} shows a histogram of
all locked mode phase data, normalized by mode duration and total number of modes in the given set: each mode
contributes a total of $100/N$, where $N$ is the total number of disruptive or non-disruptive modes; and further normalized by the binsize of 5 degrees. 
A clear $n=1$ distribution is observed in figure \ref{fig:PhaseDistribution}a (left-hand helicity discharges), with a peak at
$\sim125^\circ$ for disruptive modes and $\sim110^\circ$ for non-disruptive modes. Figure \ref{fig:PhaseDistribution}b shows right-hand helicity discharges. The disruptive distribution shows an $n=1$ component, though an $n=2$ component is also clearly visible. The non-disruptive distribution shows a strong $n=2$ component. The presence of $n=2$ might be due to occurrences of both over and under correction of the intrinsic error field. Alternatively, the $n=2$ distribution might arise from the presence of both locked and quasi-stationary modes. Locked modes are expected to align with the residual, whereas quasi-stationary modes are expected to move quickly past the residual, spending the most time in the anti-aligned phase. Similar analysis on subsets of these data shows consistent results, suggesting that these distributions are not specific to a certain experimental campaign. 

The intrinsic $n=1$ error field in DIII-D has been characterized by in vessel apparatuses \cite{LaHaye1991,Luxon2003} and the errors attributed to poloidal field coil misalignments and ellipticity, and the toroidal field buswork. It is found that the intrinsic error is well parameterized by the plasma current $I_p$ and the toroidal field $B_T$; "standard error field correction" in the DIII-D plasma control system calculates $n=1$ correction fields based on these. The majority of DIII-D plasmas are run with $I_p$ in the counter-clockwise direction and $B_T$ in the clockwise direction when viewed from above, referred to as the "normal" directions.  Taking ranges for $I_p$ and $B_T$ expected to encompass the majority of left-hand helicity discharges (i.e. $I_p = 0.8$ to 1.5 MA and $B_T = -1$ to -2.1 T), the standard error field correction algorithm applies correction fields between $-164^\circ$ and $-110^\circ$. The preferential locking angles shown in figure \ref{fig:PhaseDistribution}a might be due to a \textit{residual} EF, resulting from the vector addition of the intrinsic EF and an imperfect correction field. 

Figures \ref{fig:PhaseDistribution}c-d show how residual fields can arise from changing intrinsic and/or correction fields. Figures \ref{fig:PhaseDistribution}e-f illustrate how the distributions in \ref{fig:PhaseDistribution}a-b might look if the residual is reproducible, or not. In the former case, a narrow, peaked distribution is expected, whereas in the latter case, a broad, flat distribution is expected. This might explain the distributions in figures \ref{fig:PhaseDistribution}a-b. In addition, quasi-stationary modes are also present in these data, and contribute to the broadening.

\subsection{Change in $n=1$ field at locking and growth before disruption}
\label{sec:disruptiveGrowth}

\begin{figure}[h]
        \includegraphics[trim=1cm 6.2cm 1cm 6.5cm, clip=true, scale=0.45]{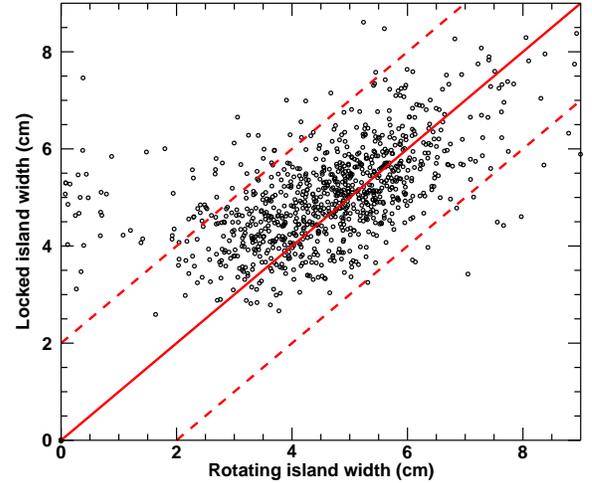} 
\caption{The
        mode width before and after locking, as calculated from the Mirnov
        probe array and ESLD measurements respectively. The rotating mode
        width is evaluated when mode rotation reaches 2 kHz; locked mode
        width evaluated at 50 ms after locking 
        to allow decay of shielding currents in the wall. The solid line shows where the widths are equal, while the dashed red lines quantify the conservative $\pm2$ cm error bar on 
the island width estimates.}  \label{fig:ModeAmpBeforeAndAfterLocking}
\end{figure}

Figure \ref{fig:ModeAmpBeforeAndAfterLocking} shows the change in mode width as the NTM slows from rotating at $f=2$ kHz to
50 ms after locking. To within the $\pm 2$ cm errors on the width estimate from magnetics, a significant growth or decay at locking is not observed for the majority of modes.
Only $\sim$10\% of locked modes appear above the top diagonal dashed line, indicating a growth at locking beyond error.

\begin{figure*}
        \centering \includegraphics[scale=0.45]{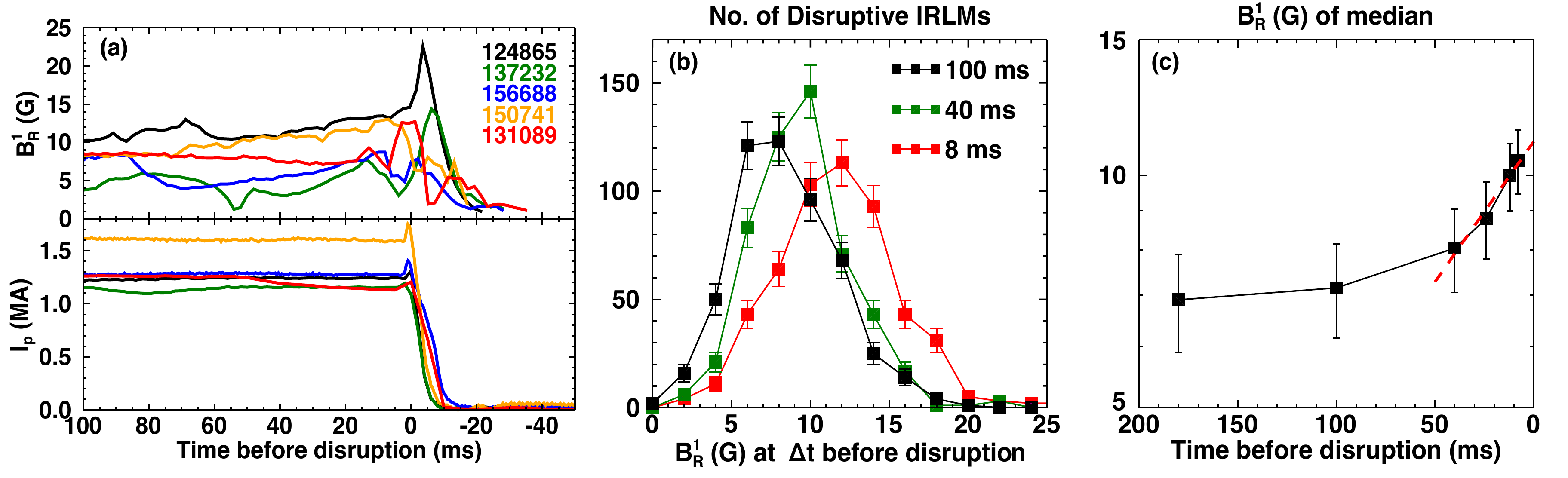} 
\caption{
	(a) $n=1$ radial field and plasma current traces from five randomly chosen disruptive IRLMs. 
	(b) Histograms of $B_R$ for all disruptive IRLMs at times approaching disruption. 
	(c) The medians of histograms at six time-slices (three shown in (b), and three not shown) undergo growth consistent 
	with exponential within 50 ms of the disruption, as shown by the linear trend on the semi-log plot. From the slope of the dashed line,
	an $e$-folding time in the range $\tau_g=[80,250]$ ms is estimated.
        } \label{fig:pub_GrowthRates}
\end{figure*}

It has been observed across the database that the 
radial field $B_R$ measured by the ESLDs tends to grow before
disruption. This growth is distinct from the dynamics of the locking
process, as it often occurs hundreds of
milliseconds after locking. Figure \ref{fig:pub_GrowthRates}a shows
the $B_R$ behavior before disruption for five randomly chosen
disruptive IRLMs. A general period of growth occurs between $\sim100$
and 5 ms prior to the disruption, followed by a sharp rise in $B_R$
within milliseconds of the current quench, marked by the transient rise in
$I_p$. Although interesting in its own
right, we will not investigate the details of the $B_R$ spike occurring
near the current quench, as we are interested in the dynamics leading up to
the thermal quench.

Histograms in radial field $B_R$ are plotted in 
figure \ref{fig:pub_GrowthRates}b, at times before disruption,  
to study the average evolution. The $B_R$ field in a single case often follows a complicated trajectory, as evidenced by figure \ref{fig:pub_GrowthRates}a, but
these histograms reveal global trends. In Figure \ref{fig:pub_GrowthRates}b, the centers of the 
distributions shift to higher values of $B_R$ as the disruption is approached. The median is chosen to be the representative point in the analysis that follows. The
median is the point on the distribution which divides the area under
the curve equally. 

The median is plotted in 
figure \ref{fig:pub_GrowthRates}c as a function of time (note the vertical axis is logarithmic). 
The median major radial field is about 7 G at saturation (i.e. 50 ms after locking). Later in the lifetime of the mode, and approximately 50 ms prior to the disruption, the median grows consistent with an exponential from the saturated value, reaching a final value of $> 11$ G at $\sim5$ ms before the disruption. This time is approximately coincident with the onset of the thermal quench. From the slope of the dashed line, in figure \ref{fig:pub_GrowthRates}c, an exponential $e$-folding time for $B_R$ in the range $\tau_g=[80,250]$ ms is found.  

\begin{figure}
        \centering \includegraphics[scale=0.45]{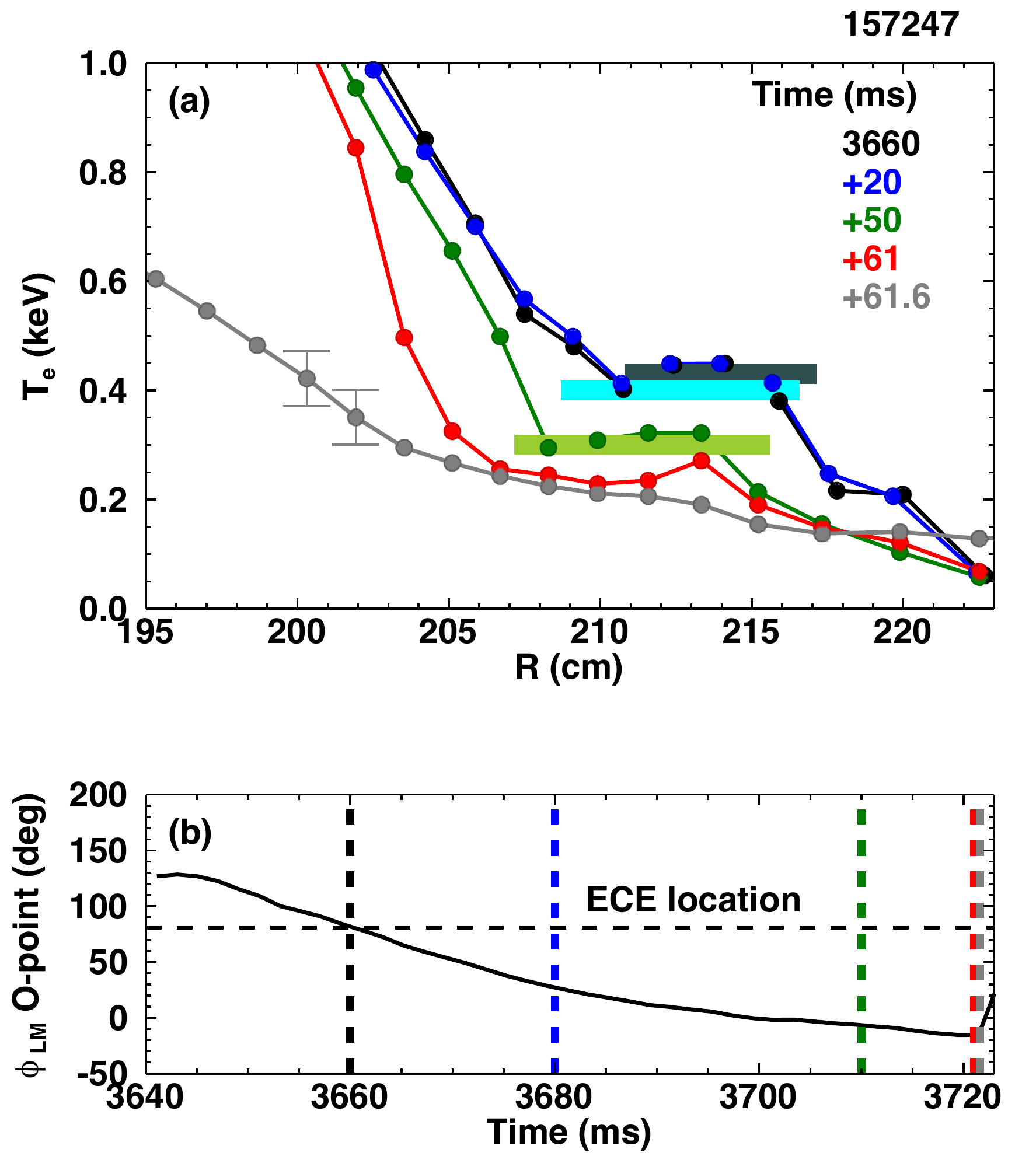} 
\caption{
	(a) Electron temperature $T_e$ profiles from ECE prior to an IRLM disruption show a clear flattening at the $q=2$ surface. Horizontal bars show
	the automated estimation of island position and width (note the vertical position of the bars is chosen for visual purposes only). 
	Two error bars are shown on the gray profile, and are representative of all $T_e$ measurement errors. 
	(b) The toroidal position of the island O-point
	on the outboard side in the midplane. The horizontal dashed line shows the toroidal location of the ECE diagnostic, and the vertical dashed lines show 
	the time slices from (a). 
        } \label{fig:ECE}
\end{figure}

The present analysis is not sufficient to discern between possible
sources of this increased $B_R$. Out of the 26 IRLMs for which 2/1 modes can be
clearly seen on ECE (see discussion in
section \ref{sec:detectLocked}), less than 10 of these provide a view
of the island O-point during the disruption. One such case is shot 157247 shown in
figure \ref{fig:ECE} which disrupted at $3723$ ms. A
significant flattening of the $T_e$ profile is seen at each time
slice, confirming that a 2/1 island with $w>5$ cm is
present. The solid horizontal bars show the
calculated island position (from EFIT) and width (derived from $\delta
I$, see Appendix B). The solid line in figure \ref{fig:ECE}b shows the toroidal location
of the island O-point on the outboard side in the midplane, and it is seen to intersect the location of the ECE diagnostic
(dashed horizontal) at $t\approx3660$ ms. The
vertical dashed lines show the times of the profiles in figure \ref{fig:ECE}a. Note that the worst toroidal alignment for O-point viewing occurs for
$\phi_{LM} = -99^\circ$, where the X-point is aligned with the
ECE. The horizontal bar at 3710 ms (green in online version) looks to
over predict the island width by $\sim3$ cm, though the island
toroidal alignment with the ECE is intermediate between the O and X-points, where the flattened region is expected to be smaller. EFIT data do not exist for the last two timeslices.

In all shots where magnetics predict O-point alignment with ECE during the
disruption, a clear flattening of the $T_e$
profile like that shown in figure \ref{fig:ECE} is observed. We
conclude that the assumption of the presence of an $m=2$ island during
the disruption is accurate, and in the shots where ECE data are available, the predicted island widths are
reasonable.

If we assume that the increase in the $n=1$ field
measured by the ESLDs is due to growth of the 2/1 island, we can estimate how much the island width would
increase. We assume a constant $B_T$, $R$, and $dq/dr|_{q2}$ during the
$\sim100$ ms of growth, where the field increases from 7 to 11 G. 
Therefore, from equation \ref{eq:cylWidth}, we expect the island width to
increase by $\sqrt{11/7}\approx 1.25$. With an average saturated width
for disruptive islands of $\sim4-6$ cm (this will be shown in
section \ref{sec:rhoq2}), a disruptive island would be expected to
grow $\sim1-1.5$ cm during the exponential growth. With a
channel spacing of 1-2 cm for ECE measurements, and with less than 10
disruptive IRLMs whose O-points are well aligned with the ECE
diagnostic during this time interval, validating this small change in
island width would be difficult. 

As the poloidal harmonic of this exponentially growing $n=1$ field is
unknown, in principle it is possible that the 2/1 island is unchanging, while
an $m=1$ or $m=3$ instability is appearing and growing. Coupling of
the 2/1 and 3/1 modes has been observed on
ASDEX-Upgrade \cite{zohm} and investigated in numerical
studies \cite{Igochine}, while coupling of the 2/1 and $m/n=1/1$
has been observed on TEXTOR \cite{deVries95}, the RTP
Tokamak \cite{schuller}, and studied in
simulations \cite{Sykes}. Although $m=4$ might be a candidate for this
growth in some shots, it will be shown in section \ref{sec:rhoq2} that
more than half of the disruptive discharges have $q_{95}<4$, so it
could only explain a minority of cases. Similarly, out of 13 IRLMs
that occur in plasmas with $q_{95}<3$, at least 7 show a clear
disruptive growth, in which the $m=2$ or $m=1$ modes are the only
candidates for this growth. 


Note that although we have chosen to characterize the final disruptive growth prior to the final thermal quench here, a single mode might undergo multiple growths and minor disruptions, as shown in figure \ref{fig:eigspecExample}.

\section{Interdependence of locked modes and $\beta$}
\label{sec:beta}

\subsection{Effect of locked mode on $\beta$ and equilibrium}
\label{sec:betaDecrease}

A common sign of the existence of a LM is a reduction in plasma
$\beta$. Here we investigate $\beta_N = \beta (aB/I_p)$ as it
has been shown to affect NTM onset thresholds \cite{Buttery2008} ($a$ is the plasma minor radius and $B$ is usually taken to be the toroidal field on axis). 

\begin{figure*}[t]
	 \includegraphics[scale=0.21]{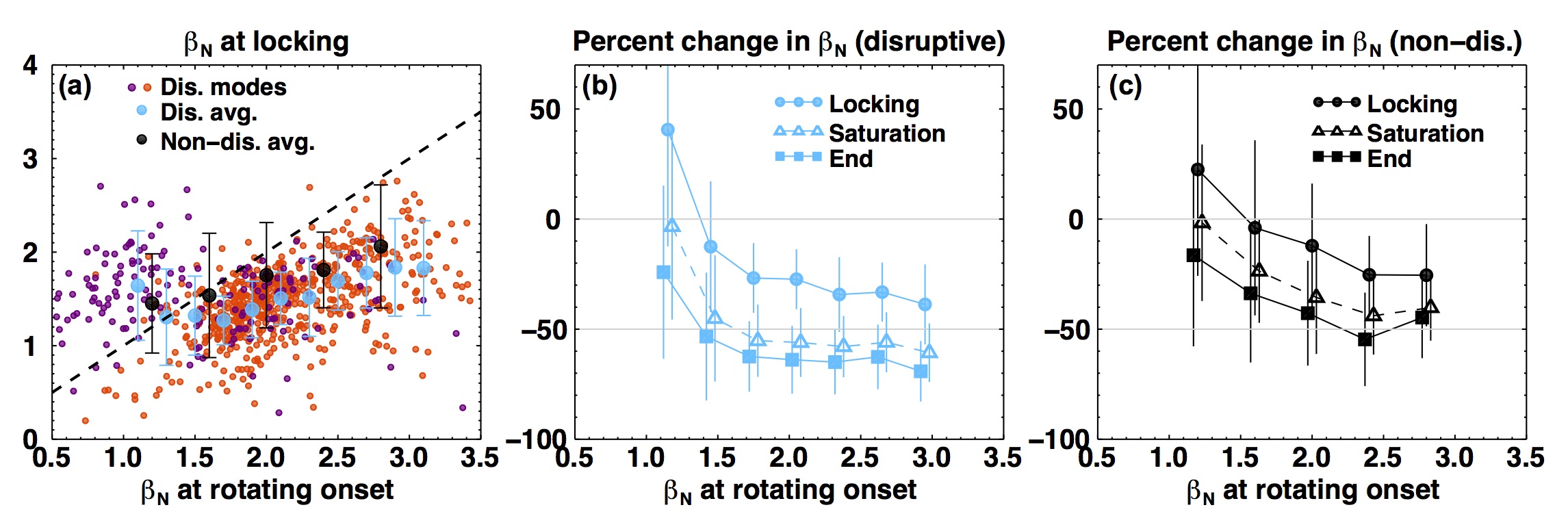}
 \caption{
 	(a) The time evolution of $\beta_N$ between onset of the rotating mode and mode locking. The blue
	and black points show the average and standard deviation of the disruptive (red and purple) and 
	non-disruptive (raw data not shown) populations. The purple disruptive IRLMs are preceded by another LM (red are not).
        (b) The percent change in $\beta_N$ during each phase of disruptive IRLMs, as compared with the $\beta_N$ at time of rotating onset. 
	(c) Same as (b), but for non-disruptive IRLMs. 
}  
        \label{fig:satBeta_peakBeta_Disruptive}
\end{figure*}

Figure \ref{fig:satBeta_peakBeta_Disruptive}a
shows $\beta_N$ at locking as a function of $\beta_N$ at mode onset. Raw data for
disruptive modes are shown in red and purple. The purple disruptive
modes are preceded by an earlier LM,  while the red are not. The majority of the red
points are observed to lie below the dashed diagonal. This observation is
reiterated by the light blue points, showing the average and standard
deviation also lying predominantly below the diagonal, meaning that $\beta_N$ decreases from NTM onset to locking. 

Notice that  a significant population of purple
points lie above the dashed line for  $\beta_N$ at rotating onset <
1.5. The rotating phase of these IRLMs begins with frequency $f=0$
(locked), at which point they spin up. In these cases, $\beta_N$ has suffered a large
degradation from the previous locked mode. In most of these cases, the $\beta_N$ at the following locking
is greater than the degraded $\beta_N$ at the time of spin up, and
therefore the points lie above the dashed line.

The light blue and black averages from
figure \ref{fig:satBeta_peakBeta_Disruptive}a are replotted in
figures \ref{fig:satBeta_peakBeta_Disruptive}b 
and \ref{fig:satBeta_peakBeta_Disruptive}c in the form of percent changes 
in $\beta_N$ at each stage, relative to the $\beta_N$
at rotating onset. These plots show how $\beta_N$ changes from the rotating onset to locking, to $\beta_N$
saturation, and to mode termination. The time of $\beta_N$ saturation
is taken to be 200 ms after locking (found empirically, in approximate agreement with twice the typical DIII-D
energy confinement time, $\tau_E \approx 100$ ms), and the time of mode  termination
is taken to be 50 ms before disruption or disappearance of the mode. Except for the
initial phase with $\beta_N$ at rotating onset < 1.5,  a continuous
decrease in $\beta_N$ is observed during successive phases of the IRLM. 

On average, the disruptive modes cause a larger degradation of
$\beta_N$ than non-disruptive modes 
during all three phases. For
$\beta_N$ at rotating onset > 1.6, disruptive modes cause 20-50\%
reduction during the rotating phase, 50-70\% reduction by $\beta_N$
saturation, and 50-80\% reduction within 50 ms of the disruption. 
Of course a further, complete loss of $\beta_N$
occurs at disruption (not shown).

\begin{figure}
        \includegraphics[scale=0.43]{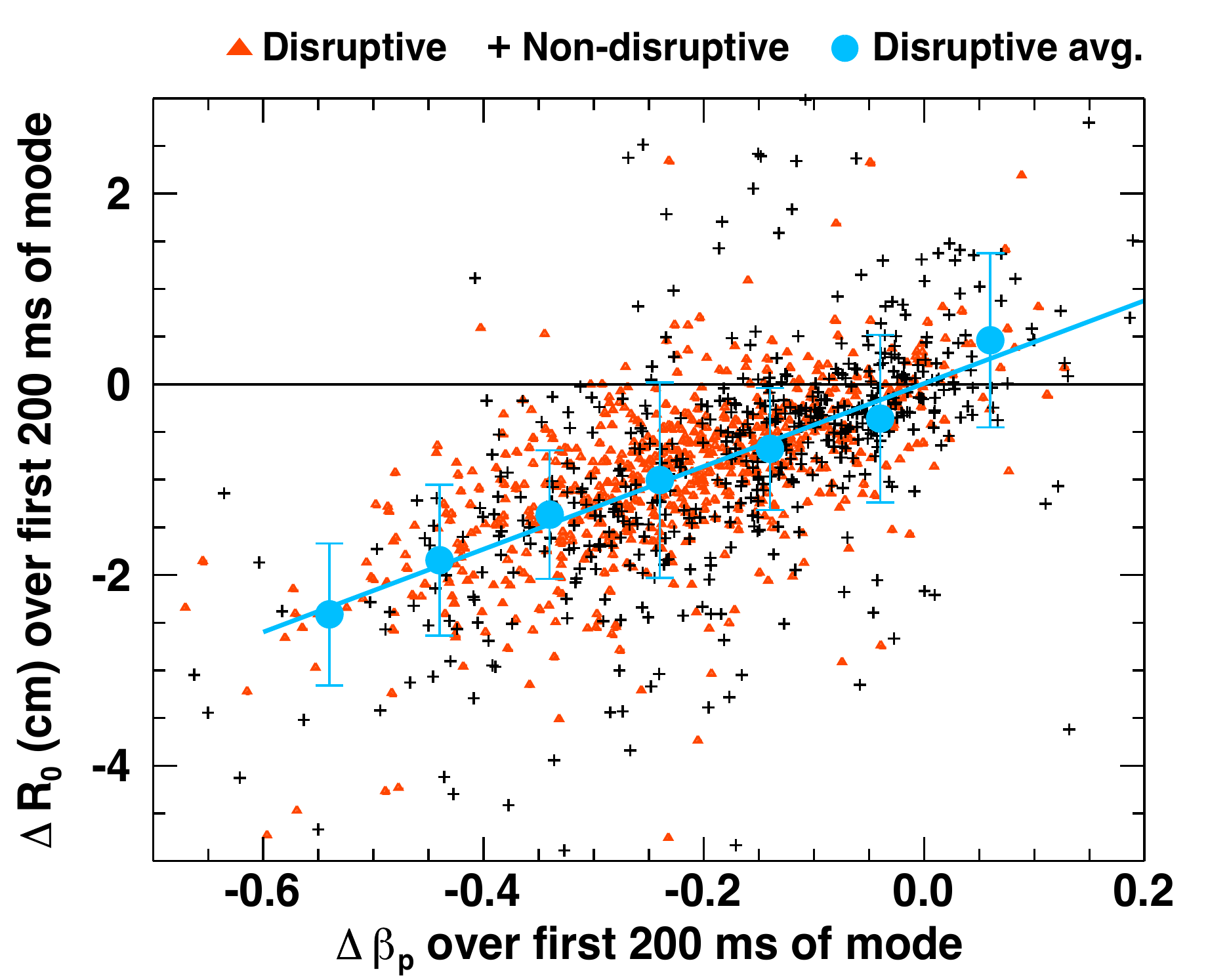} \caption{
        Both disruptive
        and non-disruptive modes show a linear dependence with $\Delta
        R_0/\Delta \beta_{p} \sim 4$ cm. A number of outliers are
        produced in the non-disruptive population by significant changes of plasma shape (e.g. diverted
        to wall-limited plasma).  } \label{fig:deltaR0_deltaBeta}
\end{figure}

With the reduction in $\beta_N$ in
figure \ref{fig:satBeta_peakBeta_Disruptive}, and assuming constant
$I_p$ and $B_T$ (which is accurate for most modes occurring during the
$I_p$ flattop), a similar reduction in $\beta_{p} = \langle p \rangle
/ (B_{\theta}^2/2 \mu_0)$ is expected as well.  The reduction in
$\beta_{p}$ causes a reduction in the Shafranov
shift. Figure \ref{fig:deltaR0_deltaBeta} shows the linear dependence
of $R_0$ on $\beta_{p}$ during the first 200 ms after locking. A linear fit
to the disruptive data provides a shift ratio of $\Delta
R_0/ \Delta \beta_{p} \approx 4$ cm. Most modes reduce the Shafranov
shift by 0-3 cm. This reduction of the $1/0$ shaping reduces toroidal coupling of the $1/1$ and $2/1$ modes, the
$2/1$ and $3/1$ modes, and other $m/n$ and $(m+1)/n$ mode combinations \cite{Fitzpatrick1994}.
The larger scatter in the non-disruptive population can be explained
at least in part by significant shape changes, such as the transition
from diverted to wall limited. For this reason, the linear fit is
performed only on the disruptive data.

\subsection{Effect of $\beta$ on the saturated width, IRLM rate of occurrence, and disruptivity}
\label{sec:betaDisruptivity}

As NTMs are the result of helical perturbations
to the bootstrap current \cite{lahaye2006}, we expect the island width
to depend on terms that drive the bootstrap current. The Modified
Rutherford Equation (MRE) has been shown in previous works to describe
3/2 \cite{butteryPPCF,lahaye98} as well as 2/1 \cite{urso,Kim2015} tearing mode saturation well. The saturated modes of
interest here have an average width $w\approx5$ cm and a standard
deviation $\sigma\approx2$ cm. Being that small island terms in the
MRE start to become small for $w>2$ cm (assuming $w_{pol}\approx2$
cm \cite{lahaye2006}), small island effects are ignored. A steady state expression of the MRE
(i.e. $dw/dt \to 0$) is given as follows,

\begin{equation}
0 =  r\Delta'(w) + \alpha \epsilon^{1/2} \frac{L_q}{L_{pe}} \beta_{p
e} \frac{r}{w}  + 4 \left( \frac{w_v}{w} \right)^2
\label{eq:Rutherford}
\end{equation}

\noindent where $\Delta'(w)$ is the classical stability index, $\alpha$ is
an \textit{ad hoc} accounting for the stabilizing effect of field
curvature ($\alpha\approx0.75$ for typical DIII-D parameters),
$\epsilon = r/R$ is the local inverse aspect ratio, $L_q = q/(dq/dr)$
is the length scale of the safety factor profile,  $L_{pe} =
-p_e/(dp_e/dr)$ is the length scale of the electron pressure profile
($L_{pe} > 0$ as defined),  $\beta_{p e}$ is the electron poloidal
beta, and $w_v$ is the vacuum island width driven by the error field
($w_v\sim1$ cm in DIII-D). Note that usually a cosine term appears on
the term involving $w_v$ \cite{FitzpatrickEF}, but it is set to unity
here, which assumes the most destabilizing alignment of the locked
mode with the error field.  

We take $\Delta'(w)$ to be of the form $\Delta'(w) =  C_0/r -
C_1w/r^2$ \cite{white}. With this definition of $\Delta'(w)$,
equation \ref{eq:Rutherford} is a nontrivial cubic equation in $w$. 
However, an approximate solution was found by approximating $(w_v/w)^2 \approx
a w_v/w -b$ (with $a=0.45$, $b=0.045$, found to be accurate to within
15\% over the domain $w_v/w = [0.12, 0.3]$). The resulting equation
is quadratic, of straightforward solution.
The approximate saturated width
expression is given by, 

\begin{equation}
\begin{array}{ll}
\frac{2w_{sat}}{r} = & \left(\frac{C_0 -4b}{C_1} \right) +  \\
& \left [ \left(\frac{C_0 -4b}{C_1} \right)^2  +\frac{4}{C_1}\left
( \alpha \epsilon^{1/2} \frac{L_q}{L_p} \beta_p  +
4a\frac{w_v}{r} \right)  \right ]^{1/2}\\
\end{array}
\label{eq:satWidth}
\end{equation}

\begin{figure}
        \includegraphics[scale=0.48]{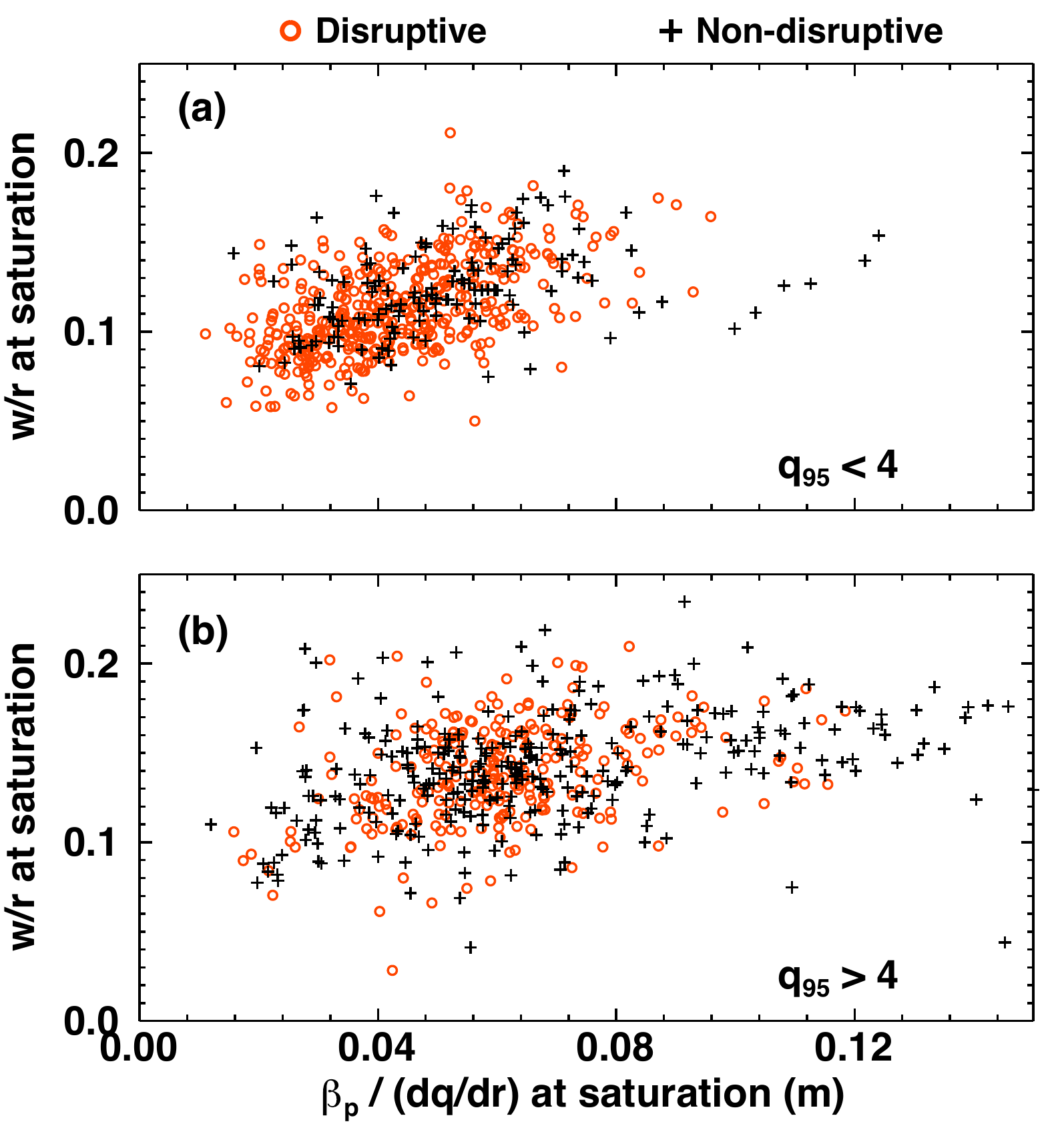} \caption{
	(a) The normalized island width at saturation as a function of $\beta_p/(dq/dr)$ for discharges
	with $q_{95}<4$. (b) Same as (a) for discharges with $q_{95}>4$.  Only islands with $w<9$ cm are
	shown here.}  
\label{fig:saturatedWidth}
\end{figure}

The data in figure \ref{fig:saturatedWidth} show the normalized island
width as a function of $\beta_p/(dq/dr)$. The parameters $L_p$ and
$\Delta'(w)$ are difficult to acquire with automated analysis, and
therefore are not available in the database. Therefore, a direct
fitting of equation \ref{eq:satWidth} is not appropriate, but the
conclusion that $w_{sat}/r$ increases with $\beta_p/(dq/dr)$ is
evident. A
correlation of $r_c=0.55$ is found between $w_{sat}/r$ and
$\beta_{p}/(dq/dr)$ for plasmas with $q_{95}<4$. Plasmas with
$q_{95}>4$ show a weaker correlation with $r_c=0.36$. This correlation suggests that locked
modes in DIII-D are driven at least in part by the bootstrap current.

\begin{figure}[t]
\includegraphics[trim=0cm 0.2cm 0cm 0cm, clip=true, scale=0.45]{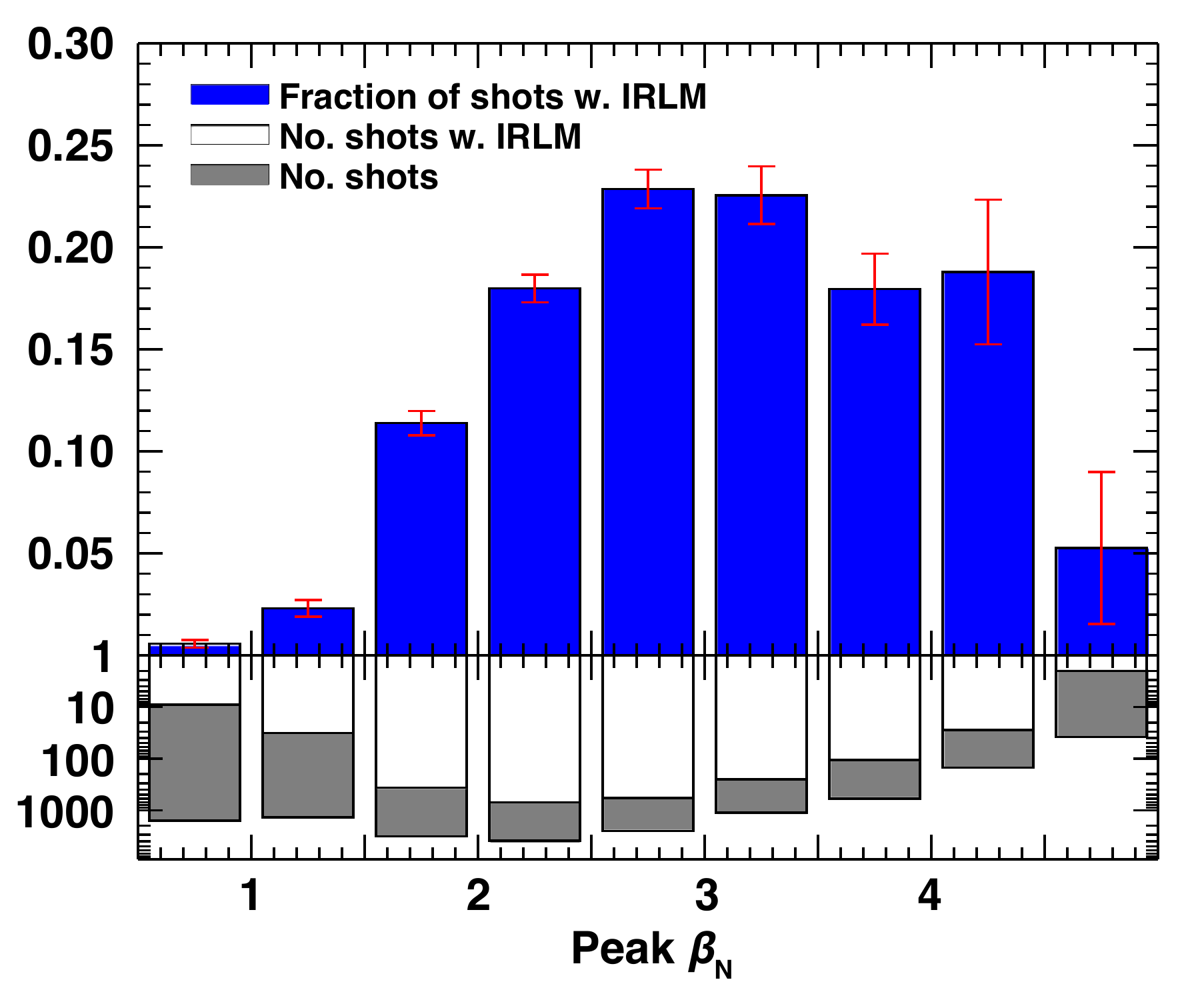}      
\caption{The shot-wise rate of occurrence of IRLMs (number of shots with IRLMs /
        total number of shots) as a function of the maximum $\beta_N$
        achieved during the shot. Blue bars (on top) are formed by the
        quotient of the white and gray (on bottom). Note the logarithmic axis for
        the lower axis.  } 
\label{fig:occurrences_disruptivity}
\end{figure}

A one dimensional study of IRLM frequency (or prevalence) versus
$\beta_N$ appears to suggest that intermediate $\beta_N$ shots are the most
prone to IRLMs, as shown in
figure \ref{fig:occurrences_disruptivity}. The fraction of shots
containing an IRLM increases with $\beta_N$ up to 2.75. 

The fraction of shots with IRLMs decreases significantly for
$\beta_N \geq 4.5$. A manual investigation of the 38 shots in the $\beta_N=4.75$ bin
reveals that these discharges often have (1) $q_{95}\geq 5$, or (2) high
neutral beam torques of $T\approx6$ Nm, or both. 
In both cases, a low occurrence of IRLMs might be expected due to a weak wall torque (see equation \ref{eq:wallTorque}): (1) 
the $q=2$ surface is far from the wall in discharges with high $q_{95}$, and 
(2) shots with high injected torque likely exhibit high plasma rotation, where the wall torque goes like
$\omega^{-1}$, and is therefore relatively small. A three-dimensional analysis of IRLM frequency
versus $\beta_N$, $\rho_{q2}$, and NBI torque might be more
informative. These data are not populated for all shots in the
database, so this analysis is reserved for future work.

\begin{figure*}[t]
\includegraphics[scale=0.9]{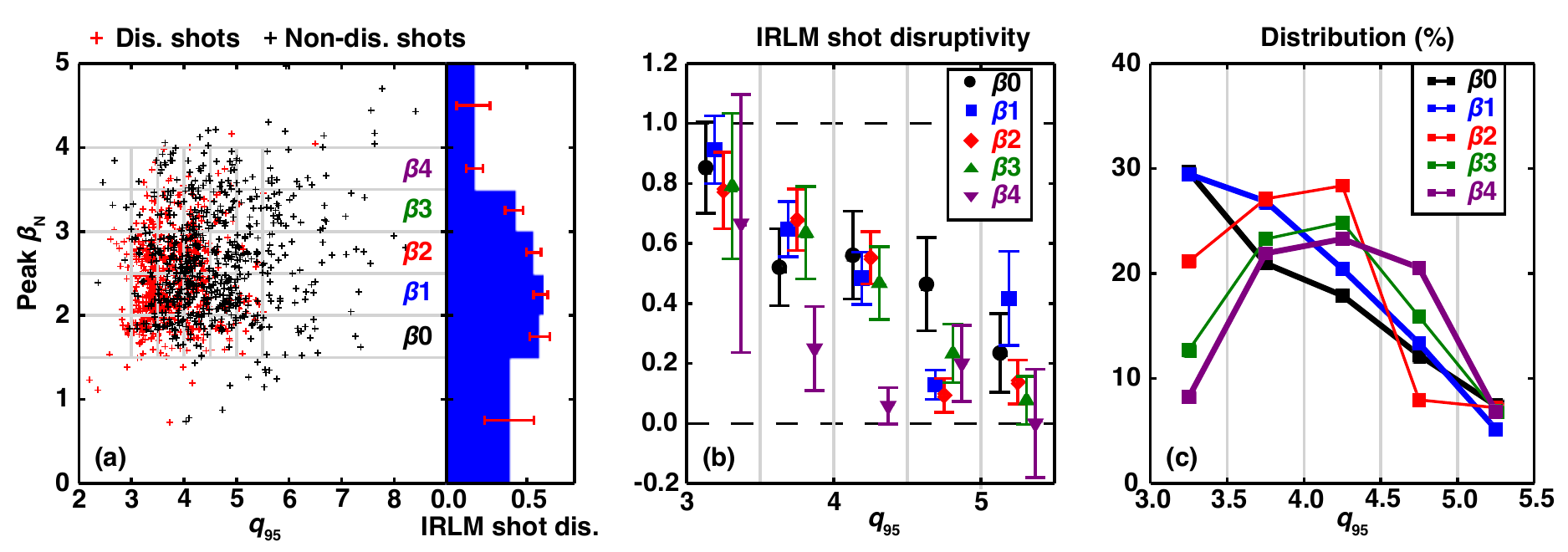}      
\caption{ 
(a) The raw data for the highest achieved $\beta_N$ as a function of
$q_{95}$. The histogram on the
right shows the one-dimensional \textit{IRLM shot disruptivity} as a function of peak
$\beta_N$. Windows in $\beta$ are labeled, and the binning for figures
(b) and (c) are shown in gray.  
(b) IRLM shot disruptivity as a function of $q_{95}$ for the binned data in (a).  
(c) The distribution of each $\beta_N$ bin in $q_{95}$ in percent.                     
} 
\label{fig:disruptivity_panel}
\end{figure*}

The \textit{IRLM shot disruptivity} as a function of $\beta_N$ and $q_{95}$ is
plotted in figure \ref{fig:disruptivity_panel}. The IRLM shot disruptivity
is defined as the "number of shots with disruptive IRLMs" divided by
the "number of shots with IRLMs". Considering IRLM shot disruptivity as a function of $\beta_N$ only,
the histogram at the right of figure \ref{fig:disruptivity_panel}a
shows a peaked distribution, with the highest values for
$\beta_N \sim 1.5-3$. Similar results were obtained for the global
disruptivity as a function of $\beta_N$ at NSTX \cite{gerhardt} (the NSTX study is
not limited to locked modes, and disruptivity is normalized by the amount of time
spent at the given $\beta_N$ value). Reduced locked mode
disruptivity at high $\beta_N$ was also observed at MAST \cite{butteryMAST}. 

The IRLM shot disruptivity dependence on $\beta_N$ can be explained in part
by the reduced number of discharges with $q_{95}<3.5$ when
$\beta_N$ is high. To see this, we bin the data in
figure \ref{fig:disruptivity_panel}a into five $\beta_N$ windows,
denoted $\beta 0$, $\beta 1$, $\dots \beta 4$. First, note that
figure \ref{fig:disruptivity_panel}b shows a general decrease in IRLM
disruptivity as $q_{95}$ is increased, across all  $\beta_N$
windows. Next, figure \ref{fig:disruptivity_panel}c shows the percent
distribution of $q_{95}$ values in each of these $\beta_N$
windows. The two highest values of $\beta_N$ (purple and green) have
the lowest percentage of discharges with $q_{95}<3.5$, and the highest
percentage of discharges with $q_{95}$ between 4.5 and 5. This
distribution of $q_{95}$ reduces the number of
disruptive discharges, with the net result that the higher $\beta_N$ discharges
appear less disruptive. 

As $q_{95}$ is shown here to affect IRLM shot disruptivity, 
fixing $q_{95}$ removes one variable, and allows us to more closely
inspect the dependence of IRLM shot disruptivity on
$\beta_N$. Figure \ref{fig:disruptivity_panel}b shows IRLM
shot disruptivity across five windows in $\beta_N$ (denoted by color), for five values of $q_{95}$ (separated by
vertical gray lines). In all $q_{95}$ windows, there is either no
significant dependence on $\beta_N$, or a lower IRLM shot disruptivity at higher $\beta_N$, particularly in the window $3.5<q_{95}<4.5$. 
Although larger islands are expected at higher $\beta_N$, it will be shown
in the next section that \textit{IRLM disruptivity} depends very weakly on
island width (refer to figure \ref{fig:quantDis}), 
which in turn might explain the weak dependence on $\beta_N$.
It should be noted that only $q_{95}$ is fixed here. Other parameters that
are correlated with $\beta_N$ (e.g. NBI torque, ion and electron temperatures,
and possibly others) are not fixed, and might affect the apparent IRLM
disruptivity scaling.

\section{IRLM disruptivity}
\label{sec:rhoq2}

\subsection{Decoupling the effects of $\rho_{q2}$, $q_{95}$, and $l_i$ on IRLM disruptivity}

\begin{figure*}
           \includegraphics[scale=0.9]{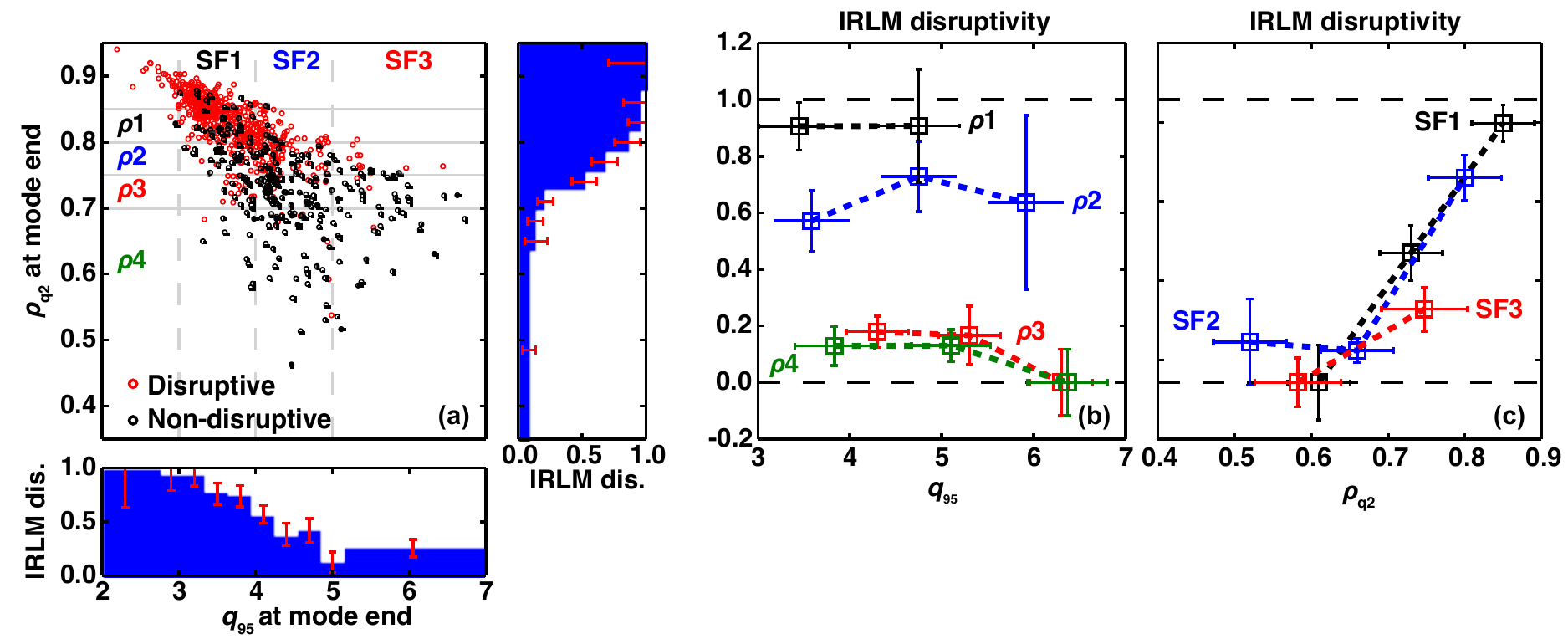} 
           \caption{
	(a) The relationship of disruptive and non-disruptive IRLMs in
	 $q_{95}$ and $\rho_{q2}$  space. One-dimensional IRLM
	 disruptivity histograms in $q_{95}$ and $\rho_{q2}$ are shown on the
	 bottom and right.  Bins in $\rho_{q2}$ and safety factor are shown in
	 gray, for use in (b) and (c). Note that only the binning intervals
	 are shown, and not the explicit bins (i.e. the dashed and solid gray
	 lines are unrelated).  
	 (b) IRLM disruptivity in windows of $\rho_{q2}$ as a function of $q_{95}$.  
	 (c) IRLM disruptivity in windows of $q_{95}$ as a function of $\rho_{q2}$.	
        } \label{fig:q95RhoCompare}
\end{figure*}

Figure \ref{fig:q95RhoCompare}a shows the normalized radius of the
$q=2$ surface $\rho_{q2}$ and $q_{95}$ at the mode end, defined here
to be 100 ms prior to the termination of the mode. The data are from equilibrium reconstructions
constrained by both the external magnetics, and the Motional Stark
Effect (MSE) diagnostic. On the right and below
figure \ref{fig:q95RhoCompare}a are histograms of IRLM disruptivity as
a function of $\rho_{q2}$ and $q_{95}$ respectively. The $q_{95}$
histogram shows the expected result that lower $q_{95}$ is more
disruptive, as was also seen in figure \ref{fig:disruptivity_panel}b. The total disruptivity (i.e. not limited to IRLM disruptions) is observed to increase as $q_{95}$ is decreased in DIII-D \cite{FNSF}. This is in agreement with, but not limited to, observations from JET \cite{deVries2009}, NSTX \cite{gerhardt}, and MAST \cite{butteryMAST}. The histogram in $\rho_{q2}$ has a qualitatively similar shape, with the highest IRLM disruptivity at large $\rho_{q2}$, and the lowest at small $\rho_{q2}$. 

The raw data show an expected
correlation between $\rho_{q2}$ and $q_{95}$. In a circular
cross-section, cylindrical plasma, $q_{95}$ can be defined as follows,

\begin{equation}
q_{95} = \frac{2}{\rho_{q2}^2} \left [ 1
+ \frac{I_{out}}{I_{enc}} \right ]^{-1}
\label{eq:q_95}
\end{equation}

\noindent where $I_{enc}$ is the total toroidal current enclosed by the $q=2$
surface, and $I_{out}$ is the total toroidal current outside of the
$q=2$ surface. Note that the quotient $I_{out}/I_{enc}$ behaves
similar to the inverse plasma internal inductance $l_i^{-1}$: when
$I_{out}/I_{enc}$ is large, $l_i$ is small, and vice versa. The lack
of one-to-one relationship between $\rho_{q2}$ and $q_{95}$ in the raw
data of figure \ref{fig:q95RhoCompare}a may therefore be attributed in
part to variations in $I_{out}/I_{enc}$. Toroidal geometry and plasma
shaping may also introduce variations in the relationship between
$q_{95}$ and $\rho_{q2}$. 

Empirically, we find a high correlation ($r_c=0.87$) between
$l_i/q_{95}$ and $\rho_{q2}$, suggesting a relationship of the form, 

\begin{equation}
l_i/q_{95} = \alpha \rho_{q2} + c
\label{eq:q95RhoRelation}
\end{equation}

\noindent where $\alpha = 0.67 \pm0.01$, and $c = -0.23 \pm 0.01$. This equation suggests that $\rho_{q2}$ specifies
$l_i/q_{95}$, or vice versa. We
begin by studying the effects of $q_{95}$, $l_i$, and $\rho_{q2}$ on
IRLM disruptivity individually. Then, we investigate IRLM
disruptivity as a function of $\rho_{q2}$ and $l_i/q_{95}$, which
despite the high correlation between the two, reveals that
$l_i/q_{95}$ distinguishes disruptive IRLMs better than $\rho_{q2}$. 

First, to decouple the effect of $q_{95}$ and $\rho_{q2}$ on IRLM
disruptivity, we fix one and study the dependence on the other in figure \ref{fig:q95RhoCompare}. We
approximately fix $\rho_{q2}$ by considering only data that lie in
small windows of $\rho_{q2}$ denoted $\rho 1$, $\rho 2$, $\rho 3$, and
$\rho 4$ ($\rho 4$ covers a relatively large window in $\rho_{q2}$ as
data become sparse, but IRLM disruptivity appears constant throughout
the window). IRLM disruptivity as a function of $q_{95}$ in the
$\rho_{q2}$ windows is shown in figure \ref{fig:q95RhoCompare}b. Neither
 the trace corresponding to $\rho 1$ nor $\rho 2$ show a significant trend with
IRLM disruptivity beyond error. Both $\rho 3$
and $\rho 4$ show a possible decrease in IRLM disruptivity for $q_{95}
> 5.5$, but appear constant for $q_{95} < 5.5$. 

The same study is performed on $\rho_{q2}$ by choosing windows in
$q_{95}$ (or safety factor) denoted SF1, SF2, and SF3 (figure \ref{fig:q95RhoCompare}c). The three safety
factor windows agree within statistical error on an increasing linear trend,
with $<20\%$ IRLM disruptivity for $\rho_{q2}<0.7$, and $>80\%$ for $\rho_{q2}>0.85$. 

From equation \ref{eq:q95RhoRelation}, we see that fixing $\rho_{q2}$
and varying $q_{95}$ (as in figure \ref{fig:q95RhoCompare}b)
implies a variation of $l_i$ as well. The weak or absent trend in IRLM disruptivity as a function of $q_{95}$ might be explained
by competing effects of $q_{95}$ and $l_i$. 

Similarly, equation \ref{eq:q95RhoRelation} shows that fixing $q_{95}$
and varying $\rho_{q2}$ (as in figure \ref{fig:q95RhoCompare}c)
also implies varying $l_i$. The strong dependence of IRLM disruptivity
on $\rho_{q2}$ thus suggests that
either $\rho_{q2}$, $l_i$, or both have a strong effect on IRLM
disruptivity.

To help isolate the individual effects of $\rho_{q2}$,
$q_{95}$, and $l_i$, we compare how well they separate disruptive and
non-disruptive IRLMs single-handedly.

In order to quantify separation of the 
distributions in one-dimension, we employ the Bhattacharyya Coefficient (BC) \cite{bhattacharyya}. This 
coefficient was developed to measure the amount of
overlap between two statistical distributions, and is commonly used in
image processing, particularly for measuring overlap of color
histograms for pattern recognition and target
tracking \cite{Bhatt}. For two discrete probability distributions $p$
and $q$ parameterized by $x$, the Bhattacharyya Coefficient is given
by, 

\begin{equation}
BC(p, q) = \sum_{x \in X} \sqrt{p(x) q(x)} . 
\label{eq:Bhatt}
\end{equation}

The $BC$
metric varies over the range $0 \le BC(p,q) \le 1$, where a value of 1
indicates that $p$ and $q$ are identical and perfectly overlapping. A value of 0 implies they
are completely distinct (no overlap) 

\begin{figure*}[t]
	\centering \includegraphics[scale=0.9]{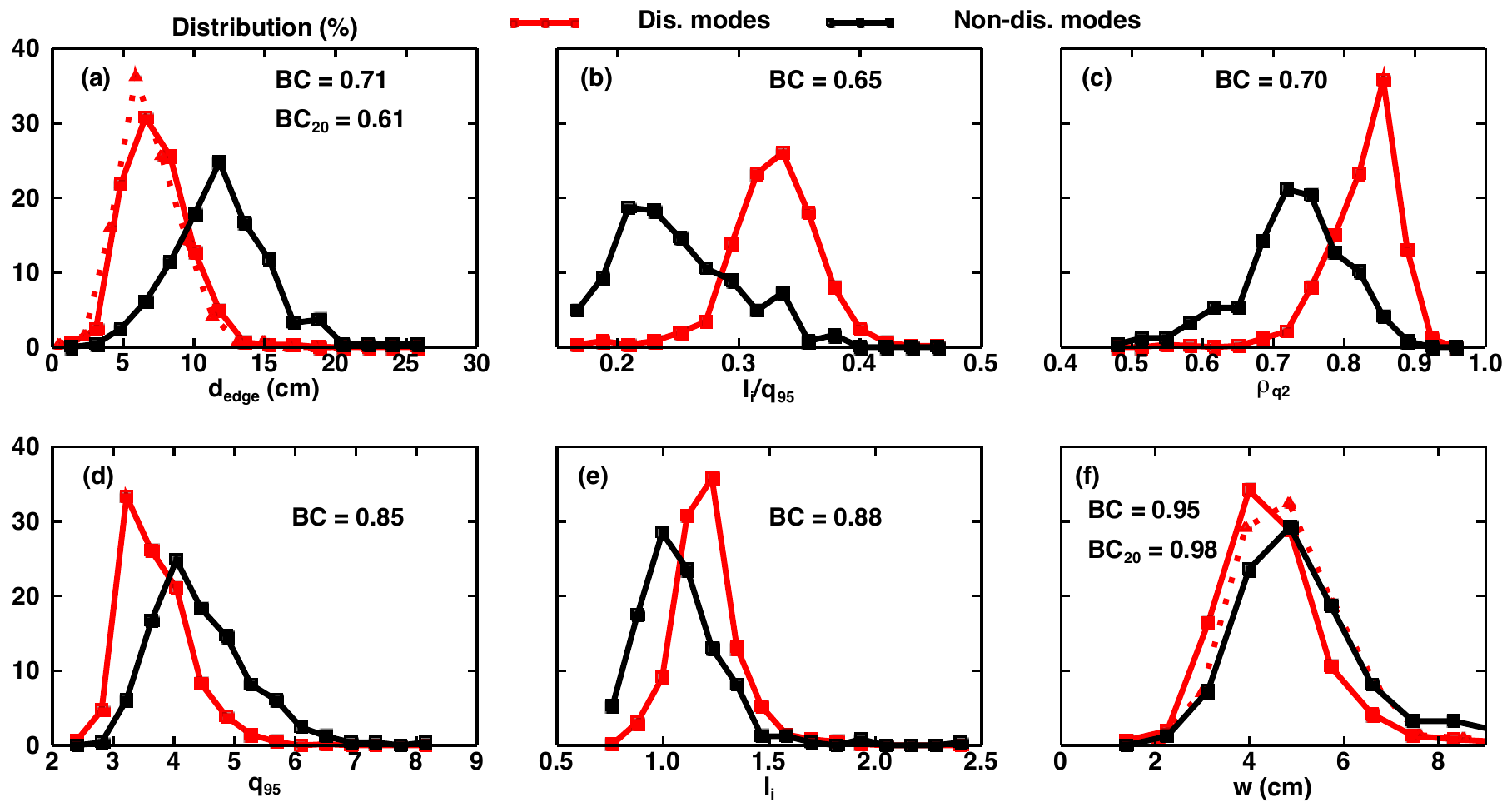}   
	\caption{
        Disruptive and non-disruptive IRLM distributions are shown as
        a function of six parameters to reveal the dominant classifier. 
        Amount of overlap  between distributions is
        quantified by the Bhattacharyya Coefficient (BC). A BC value
        of 0 indicates no overlap, and a value of 1 indicates complete overlap. 
        All BC values have an error bar of $\pm 0.04$.  
        Solid curves are evaluated at 100 ms before mode
        termination, while the dotted red in (a) and (f) are evaluated
        20 ms before the disruption, with corresponding BC$_{20}$ values 
        (the BC$_{20}$ values of all other parameters, not shown, are similar to their reported BC values).
        Not shown, $dq/dr|_{q2}$ produces a value of BC=0.89.  
        } 
\label{fig:quantDis}
\end{figure*}

Figure \ref{fig:quantDis} shows the 1D separation of disruptive and
non-disruptive modes for six parameters at 100 ms before mode
termination, and reports the $BC$ value for each. Of the three
interdependent parameters (i.e. $\rho_{q2}$, $q_{95}$, and
$l_i$), $\rho_{q2}$ is observed both visually, and by
the $BC=0.70$ value to best separate the two populations. The parameters $q_{95}$ and $l_i$ produce less separation with
coefficients of $BC=0.85$ and $BC=0.88$ respectively. All BC values reported in figure  \ref{fig:quantDis} have an 
error bar of $\pm 0.04$.

Disruptivity is found to scale strongly with plasma shaping in NSTX \cite{gerhardt}. The effects of shaping on IRLM disruptivity in DIII-D will be included in a future work.

\subsection{Decoupling effects of $\rho_{q2}$ and $l_i/q_{95}$ on IRLM disruptivity}
\label{sec:rhoStab}

\begin{figure}[t]
        	\includegraphics[scale=0.47]{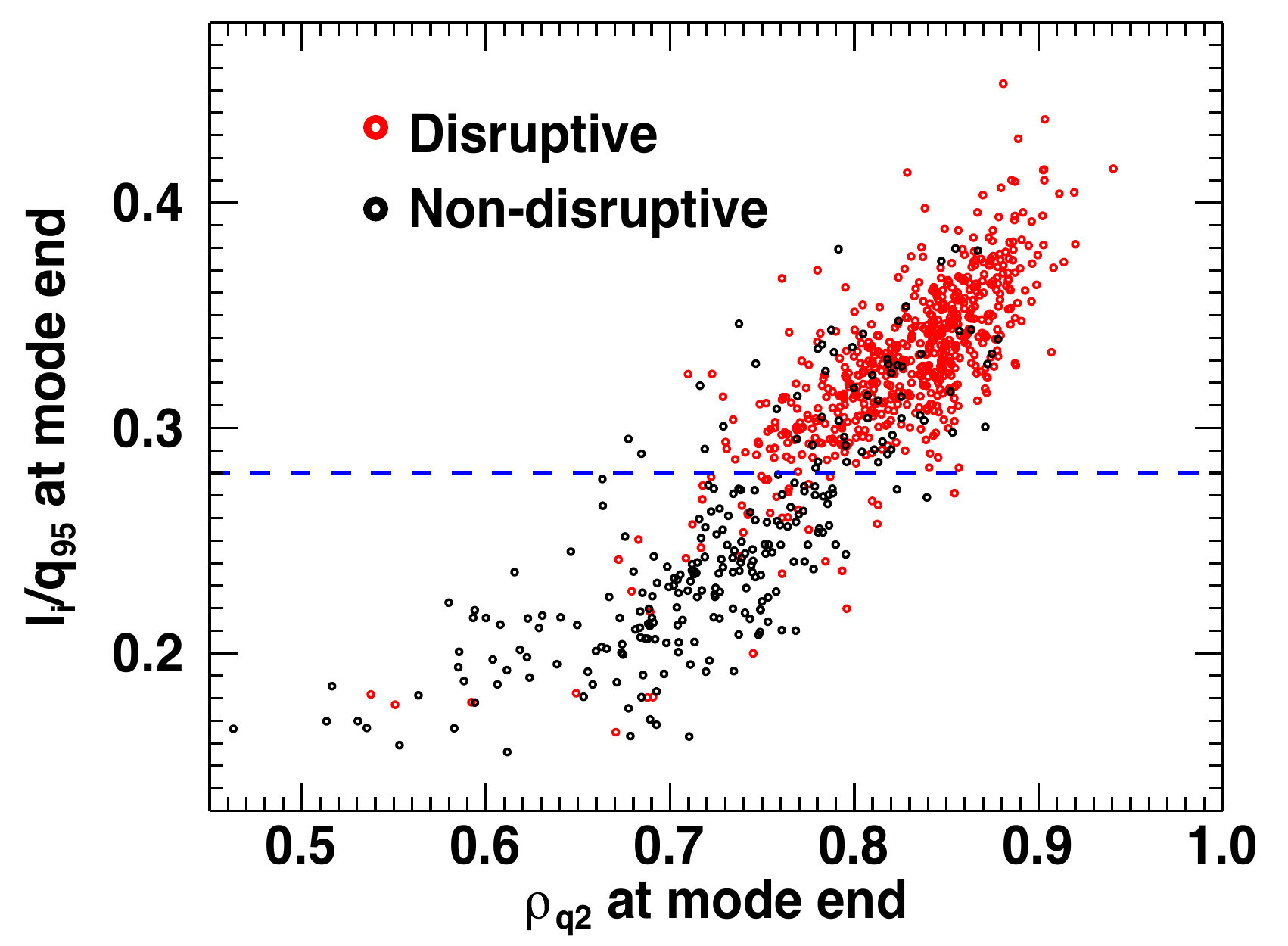}  
	\caption{
	Investigation
        	of $l_i/q_{95}$ as a function of $\rho_{q2}$ across all
        	disruptive and non-disruptive IRLMs. Mode end is 100
        	ms prior to mode termination.
	The horizontal blue
        	line at $l_i/q_{95}=0.28$ shows approximately where
        	IRLM disruptivity transitions from low to high. 
	The high
        	correlation between $l_i/q_{95}$ and $\rho_{q2}$ is
        	evident from the good clustering of the data along the
        	line  specified by
        	equation \ref{eq:q95RhoRelation}.  } 
\label{fig:stability}
\end{figure}

In the previous section, it is assumed that $l_i/q_{95}$ and
$\rho_{q2}$ are effectively equivalent, as suggested by
equation \ref{eq:q95RhoRelation}. However, here we show that $l_i/q_{95}$ has a stronger effect on IRLM disruptivity. Figure \ref{fig:stability} shows all disruptive and
non-disruptive IRLMs plotted in the 2D space of $l_i/q_{95}$ and
$\rho_{q2}$ at 100 ms prior to mode termination. 

The data in the region where
$\rho_{q2}=[0.7, 0.8]$ show a clear vertical separation, rather than
a horizontal one. For instance, choosing a threshold of
$l_i/q_{95}<0.28$ as the definition of a non-disruptive IRLM results
in 7\% mis-categorized disruptive IRLMs, and 24\% mis-categorized non-disruptive IRLMs. To capture a
similar number of correctly identified non-disruptive IRLMs using
$\rho_{q2}$, a value of $\rho_{q2}<0.78$ is used and results in 14\%
mis-categorized disruptive IRLMs, and 25\% mis-categorized non-disruptive IRLMs. This confirms the result of
figures \ref{fig:quantDis}b-c that $l_i/q_{95}$ categorizes disruptive and non-disruptive IRLMs better than $\rho_{q2}$. 

To reduce mis-categorized disruptive modes, a threshold
of $l_i/q_{95}<0.25$ can be used which produces only 3\% mis-categorized disruptive IRLMs. However, the mis-categorized non-disruptive IRLMs increase to 42\%.

Similar empirical observations of a critical boundary in $l_i$ and $q_{95}$ space are reported for density limit disruptions in JET \cite{disruptionsJET}, for current ramp-down disruptions in JT-60U \cite{Yoshino1994}, and for ``typical'' disruptions in TFTR (see figure 6 in \cite{Cheng}). Theoretical interpretations of $l_i/q_{95}$ will be presented in section \ref{sec:discussion}.

Note that $\sim17$ of the 53 disruptive discharges omitted due to lack of equilibrium data likely fall below the $l_i/q_{95}$ threshold, but are not included in figure \ref{fig:stability}, nor the categorization percentages reported in this section. The impact of these omitted discharges is assessed in section \ref{sec:prediction}.

\subsection{$d_{edge}$ discriminates disruptive IRLMs within 20 ms of disruption}
\label{sec:dedge}

In this subsection and subsection \ref{sec:widthDisruptivity}, it is assumed 
that the exponential $n=1$ growth in $B_R$ 
(figure \ref{fig:pub_GrowthRates}c) is due to growth of the
2/1 island. It will be shown here that $d_{edge}$ is as effective as
$l_i/q_{95}$ at discriminating disruptive IRLMs within 20 ms of the disruption. A possible
physical interpretation will be given in section \ref{sec:discussion}.

Figure \ref{fig:widthDisruptivity} shows the island half-width as a
function of the distance between the $q=2$ surface and the unperturbed
plasma separatrix,  $a - r_{q2}$. The shortest perpendicular distance
from the black dashed line representing the unperturbed last closed
flux surface (LCFS) to a given point is what we have called $d_{edge}
= a - (r_{q2} + w/2)$. In other words, a point appearing near the solid black line represents an island whose radial extent 
reaches near the unperturbed LCFS. Due to the perturbing field of the locked mode, we
expect kinking of the LCFS. This
kinking of the LCFS is not accounted for in our calculation of $d_{edge}$.

$d_{edge}$ is shown to separate disruptive and
non-disruptive modes as effectively as $\rho_{q2}$ at 100 ms prior to
disruption in figure \ref{fig:quantDis} (recall the $\pm0.04$ error bar on all BC values). Within 20 ms of the disruption, $d_{edge}$ separates the two
populations better with a BC value of 0.61 (compare the dotted red and solid black distributions for
$d_{edge}$ in figure \ref{fig:quantDis}a). The better separation is due to the exponential
growth of the $n=1$ field that occurs in the final 50 ms before disruption: that growth of $B_R$, and thus of $w$, decreases $d_{edge}$ without affecting $\rho_{q2}$. 
Note that evaluating $d_{edge}$ at 20 ms prior to the disruption implicitly attributes the $n=1$ exponential growth
to a growing 2/1 island, but the validity of this assumption does not affect the discrimination ability of $d_{edge}$. 

Choosing a threshold of $d_{edge}>9$ cm evaluated 20 ms before the disruption (blue dashed line in
figure \ref{fig:widthDisruptivity}), we find 8\% mis-categorized disruptive IRLMs,
and 28\% mis-categorized non-disruptive IRLMs. This is comparable with the 7\% and 24\% found for $l_i/q_{95}<0.28$, though $l_i/q_{95}$ is evaluated 80 ms earlier (these
thresholds on $d_{edge}$ and $l_i/q_{95}$ correctly categorize 195 and 194
non-disruptive IRLMs respectively). A more conservative threshold of $d_{edge}>11$
cm produces 3\% mis-categorized disruptive IRLMs, and 58\% mis-categorized non-disruptive IRLMs.

\subsection{Weak dependence of IRLM disruptivity on island width}
\label{sec:widthDisruptivity}

\begin{figure}[t]
        \includegraphics[scale=0.47]{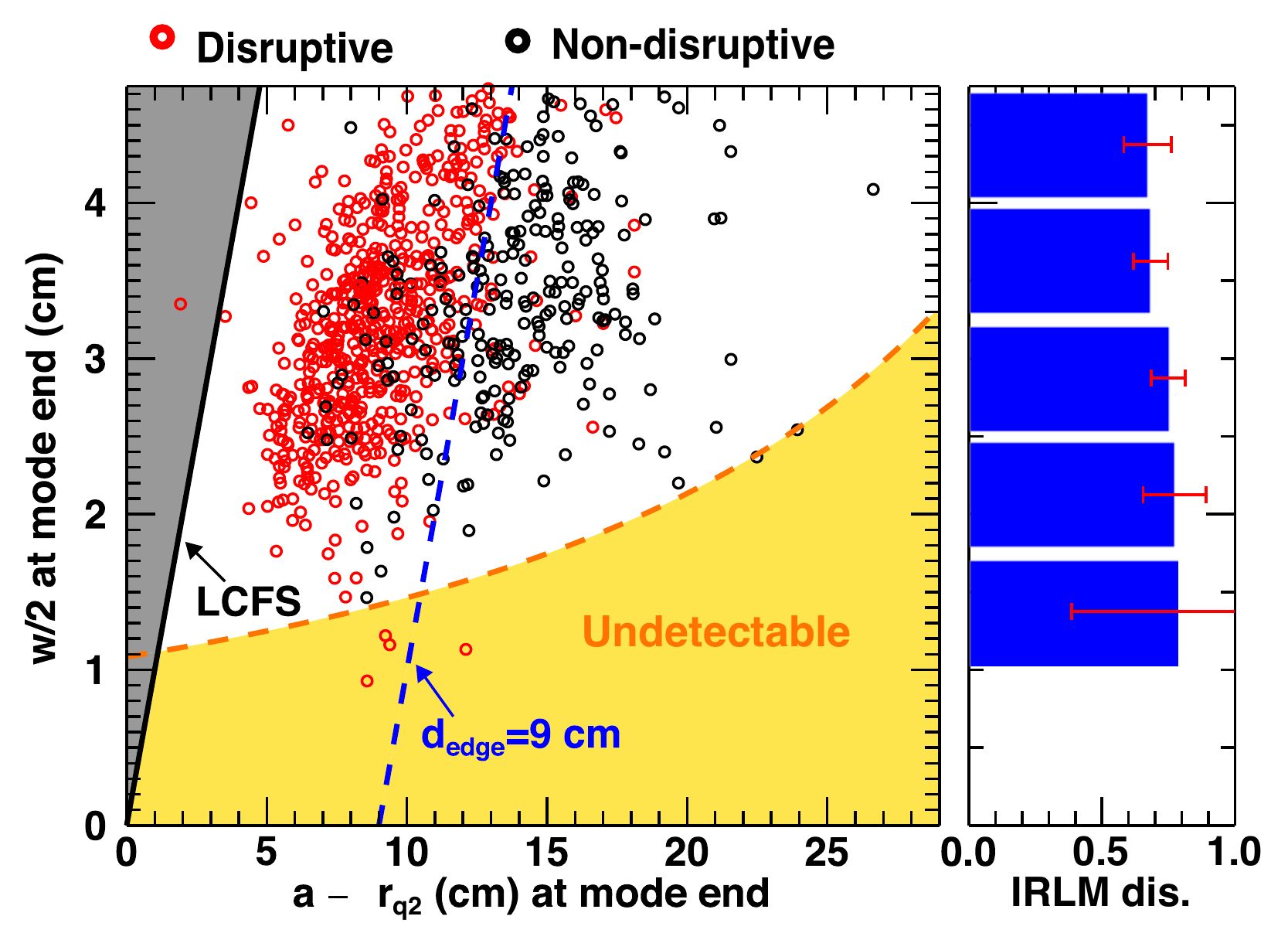} 
        \caption{
        The assumption that the 2/1 island dominates the disruptive
        exponential growth is implicit in this figure, and therefore
        this figure is exploratory. 
        The disruptive data are from 20 ms before disruption, and the 
        non-disruptive data are from 100 ms before mode termination.  The
        horizontal axis quantifies the distance of the $q=2$ surface from the
        unperturbed last closed flux surface (LCFS).  The shortest
        perpendicular distance from the unperturbed LCFS (black solid)  to a
        given point is $d_{edge}$  (i.e. $d_{edge}= a - (r_{q2} + w/2)$). The blue
        dashed line is where $d_{edge} = 9$ cm. IRLM
        disruptivity as a function of island half-width is shown in the blue
        histogram.  } 
        \label{fig:widthDisruptivity}
\end{figure}

The island width alone is found to be a poor discriminator of disruptive IRLMs
at times 20 ms before the disruption or earlier (figure \ref{fig:quantDis}f). From the time of saturation until $\sim20$ ms prior to
the disruption, disruptive and non-disruptive island widths are
similar, with a slight tendency for non-disruptive modes to be larger
at the earlier times. IRLM disruptivity as a
function of the island half-width is shown in the histogram on the
right of figure \ref{fig:widthDisruptivity}. For the majority of
islands with half-widths between 2 and 5 cm, IRLM disruptivity does
not change significantly, and might be constant within statistical
error.

The region below the curved dashed line in
figure \ref{fig:widthDisruptivity} shows the approximate mode
detection limit due to typical signal-to-noise-ratio discussed in
section \ref{sec:detectLocked}. Modes appearing in this region were
once above the detection limit, and are still measured due to the
asymmetry in onset and disappearance thresholds used in the
analysis. It is possible that undetected modes in this region might
affect the resulting IRLM disruptivity.

\subsection{IRLM disruption prediction}
\label{sec:prediction}

Thus far we have discussed percent mis-categorizations, separation of disruptive and non-disruptive distributions measured by the BC coefficient, and IRLM disruptivity all at specific points in time. Although useful for understanding the physics of IRLM disruptions, single time-slice analysis is not sufficient for disruption prediction. Prediction during the locked phase requires establishing a parameter threshold that will never be exceeded by a non-disruptive IRLM, but will be exceeded by a disruptive IRLM \textit{at some point} before the current quench.

We consider $l_i/q_{95}$ and $d_{edge}$ separately as IRLM disruption predictors. These predictors are intended to be used only in the presence of a detected locked mode. Table \ref{tab:prediction} shows the percent missed disruptions and percent false alarms for the given thresholds, and for the given warning times. The percent missed disruptions is defined as the number of disruptive IRLMs that do not exceed the threshold within $X$ ms of the disruption, divided by the total number of disruptive IRLMs. The percent false alarms is defined as the number of shots where at least one non-disruptive IRLM exceeds the threshold at any time during its lifetime, divided by all issued alarms. Note that a single non-disruptive discharge can have multiple non-disruptive IRLMs, but only one alarm. 

If all IRLMs are considered disruptive, no IRLM disruptions are missed, but the percent false alarms increases to $25\pm2\%$. This case is shown as the third row of table \ref{tab:prediction}, labeled ``none'', as no additional condition is required before issuing an alarm. Declaring an IRLM disruptive when it locks provides a warning time that depends on the survival time of the IRLM. The distribution of survival times, and therefore also of warning times for this disruption prediction criterion, are shown in figure \ref{fig:SurvivalTime}.

\begin{table}[h]
        \begin{tabular}{| l | c | c | c | c |}
        \hline
        \textbf{Condition} 	& \multicolumn{2}{l|}{\textbf{(\%) Missed}} 	& \multicolumn{2}{l|}{\textbf{(\%) False }}  \\     
 	\textbf{with IRLM} 	&  \multicolumn{2}{c|}{\textbf{disruptions}} 	&    \multicolumn{2}{c|}{\textbf{alarms}}   \\
				 	&  \multicolumn{2}{l|}{\textbf{with warning}} 	&    \multicolumn{2}{l|}{\textbf{with warning}}   \\
 				 	&  \textbf{100ms} 		& \textbf{20ms} 			&   \textbf{100ms} 	& \textbf{20ms}   \\	
        \hline
	$l_i/q_{95} >0.28$ 	&  $7\pm1$ 		& $5\pm1$			& $11\pm1$ 	& $10\pm1$ \\     
	\hline	
	$d_{edge} < 9$ cm 	&  $6\pm1$  		& $4 \pm1$ 			& $12\pm1$ 	& $12 \pm 1$ \\
	\hline
	None 			&  \multicolumn{2}{c|}{0}					& \multicolumn{2}{c|}{ $25\pm2$ }  \\
	\hline
        \end{tabular}
        \caption{IRLM disruption prediction statistics for $l_i/q_{95}$, $d_{edge}$, and no prediction parameter. The two thresholds are shown graphically by the dashed
        blue lines in figures \ref{fig:stability} and \ref{fig:widthDisruptivity} (note that figures \ref{fig:stability} and \ref{fig:widthDisruptivity}  are evaluated at single time slices, whereas
        these statistics are evaluated over appropriate time intervals). These prediction criteria are intended for use in the presence of a detected IRLM only. The "None" condition
        shows the disruption statistics for the case where all IRLMs are considered disruptive (see text for warning times for this case). 
        Note that discharges omitted by the automated analysis might increase
        the missed disruption percentages by up to 3\% (see text for details).}
        \label{tab:prediction}
\end{table}

Despite the seemingly short
warning timescale, 20 ms is a sufficient amount of time to deploy massive
gas injection in DIII-D \cite{MGI}. 

Out of the 53 disruptive discharges omitted due to insufficient equilibrium data (see section \ref{sec:detectLocked}), 17 of them might contribute to increasing the missed disruption percentages associated with the $l_i/q_{95}$ criterion reported in table \ref{tab:prediction} by up to $3\%$, with negligible effect on the percent false alarms. Also, out of the remaining $53-17=36$ disruptive discharges that satisfy $l_i/q_{95}>0.28$, 32 exhibit survival times around 20 ms, and thus the warning time for these modes is no longer than 20 ms. The increase in missed disruptions can be reduced to 1.5\% by considering discharges with $q_0<1.6$ only, as 9 of the 17 discharges contributing to the increased percent missed disruptions satisfy this criterion (only 40 discharges with 2/1 IRLMs in the entire database satisfy this criterion). The effects of these omitted discharges on the $d_{edge}$ prediction criterion are expected to be similar to those just discussed for the $l_i/q_{95}$ prediction criterion.

\subsection{$\rho_{q2}$ evolution}
\label{sec:rhoq2Evolution}

\begin{figure}[t]
	\centering
        \includegraphics[scale=0.42]{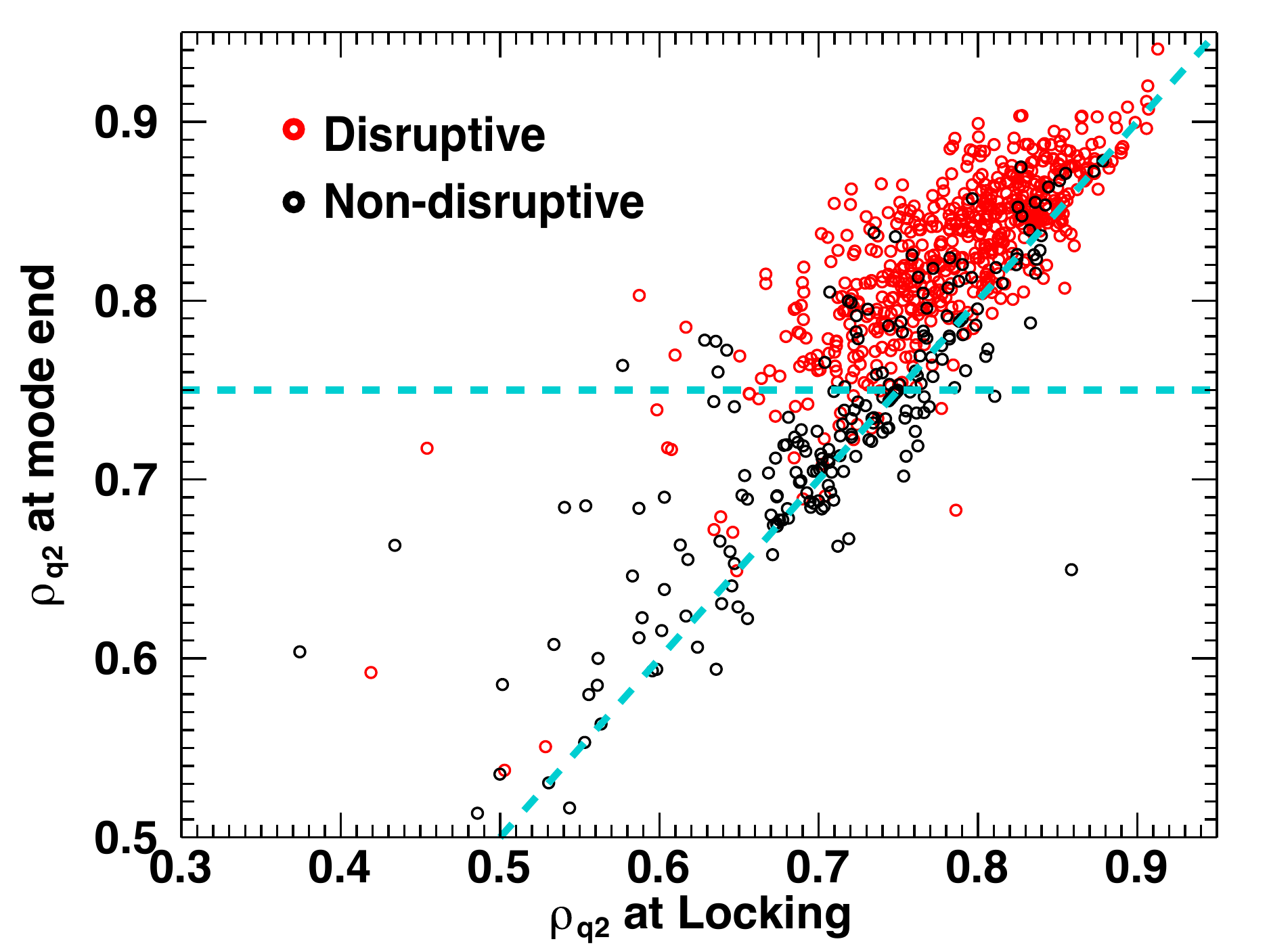}   
        \caption{
       The evolution of $\rho_{q2}$ from locking to mode end (100
       ms prior to mode termination) for IRLMs which terminate predominantly during
       the $I_p$ flat-top.  The diagonal line represents unchanging $\rho_{q2}$. The horizontal line is at $\rho_{q2}=0.75$, and marks
       an approximate transition from low to high IRLM disruptivity.
       } 
       \label{fig:rhoHist}
\end{figure}

The evolution of $\rho_{q2}$ between
locking and mode end, provides insight into the evolution of
both $l_i/q_{95}$ and $d_{edge}$. Figure \ref{fig:rhoHist}a shows that
$\rho_{q2}$ tends to increase by $\sim5\%$ between locking and mode end. 
This is noticeable by the clustering of points above the dashed diagonal. Equation \ref{eq:q95RhoRelation} suggests that an increase in
$\rho_{q2}$ at fixed $q_{95}$ implies an increase in $l_i$. 
As $I_p$, $B_T$, and plasma shaping are fixed via feedback in most DIII-D
discharges, the assumption of constant $q_{95}$ is very reasonable. 
The increase in $l_i$ during locked modes agrees with earlier works 
\cite{scoville,disruptionsJET,schuller}.

\begin{figure}[t]
        	\includegraphics[scale=0.47]{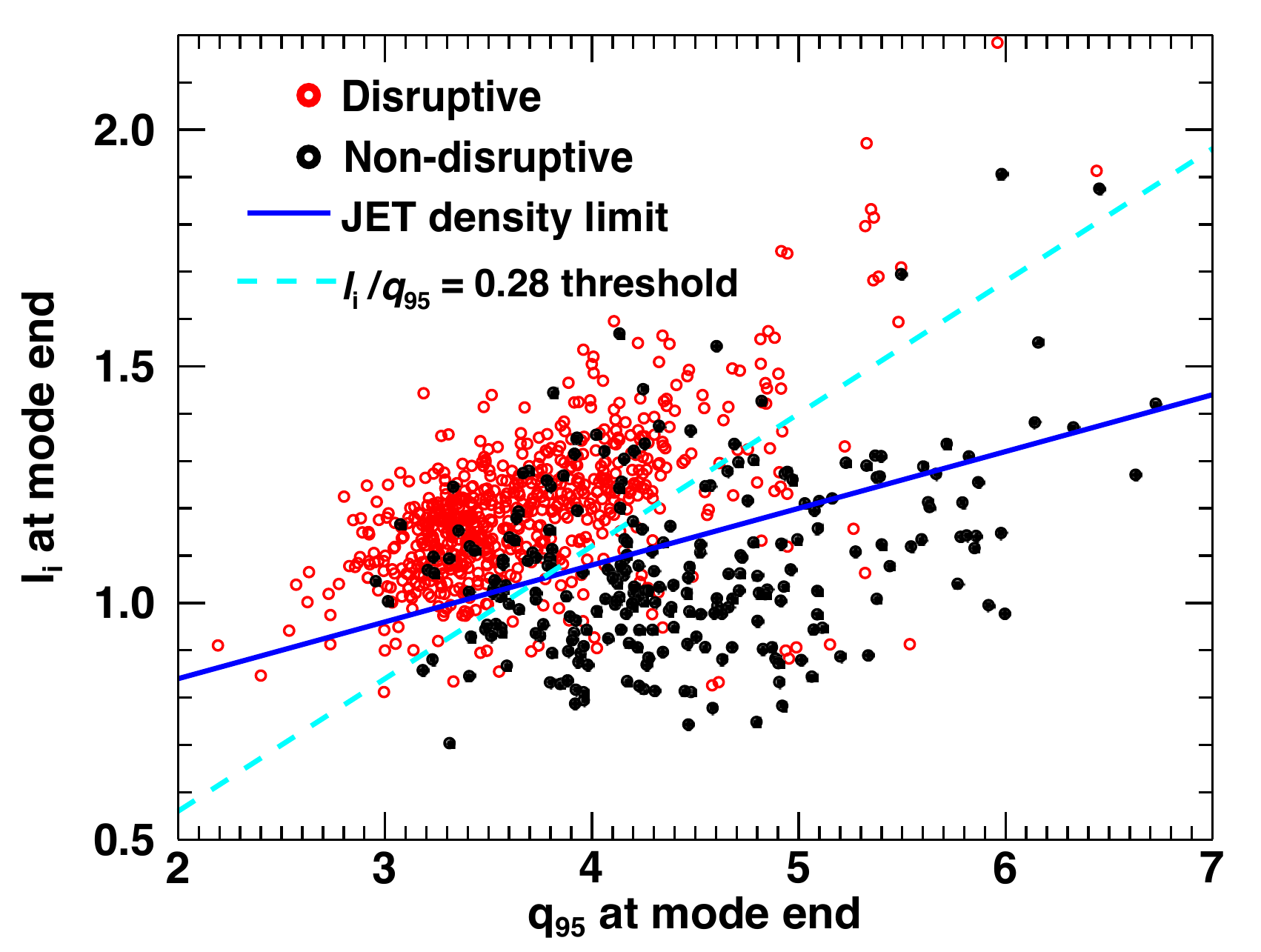}  \caption{All
        	disruptive and non-disruptive IRLMs shown in $l_i$
        	and $q_{95}$ space. The JET density limit shown in
        	blue defines the approximate lower bound of the disruptive locked
        	modes in DIII-D remarkably well. The cyan dashed line
        	is the $l_i/q_{95}=0.28$ value again, first seen in
        	figure \ref{fig:stability} to divide the populations
        	well.  } 
\label{fig:radiativeIsland}
\end{figure}

\section{Discussion}
\label{sec:discussion}

The physical significance of $l_i/q_{95}$ might be related to the potential energy available for tearing growth (i.e. to the island width dependent classical stability index $\Delta'(w)$).

A one-dimensional simulation including both MHD evolution and transport has shown that the constraint $q_{min} >1$ enforced by the sawtooth instability has a significant effect on the current profile, resulting in a steepening of the current gradient between the $q=1$ and $q=2$ surfaces as the edge $q$ value is decreased \cite{Turner1982}. A steepened current profile on the core-side of the 2/1 island is classically destabilizing, and thus, $\Delta'(w)$ is shown to increase as the edge $q$ decreases. 

A three-dimensional simulation \cite{Bondeson1986} later corroborated the classically destabilizing phenomenon found in \cite{Turner1982}. Citing these works, a study followed \cite{Cheng}, where $\Delta'(w=0)$ stability was investigated as a function of the current profile shape, with fixed axial and edge $q$ values. In that study, monotonically decreasing current profiles with $q_0\sim1$ were varied in search of a $\Delta'$ stable profile, and a limit in $l_i$ and $q_{95}$ space was found where no stable solution exists (see figure 6 of \cite{Cheng}). That limit is similar to the empirical IRLM disruption limit shown by the cyan dashed line in figure \ref{fig:radiativeIsland}. These three works together \cite{Turner1982,Bondeson1986,Cheng} provide a theoretical basis for the hypothetical connection between $l_i/q_{95}$ and $\Delta'$.

$l_i/q_{95}$ might be responsible for the pre-thermal quench growth shown in figure \ref{fig:pub_GrowthRates}, and it might also influence the evolution of MHD during and after the thermal quench.

Separately, it is interesting that at high density, a high value of $l_i$ is itself unstable, causing radiative contraction of the temperature and current profiles \cite{disruptionsJET}, which further increases $l_i$. Since $q_{95}$ is approximately fixed via feedback, increasing $l_i$ implies increasing $l_i/q_{95}$.

Alternatively, the physical significance of $l_i/q_{95}$ might be related to an excess of radiative losses in the island, overcoming the Ohmic heating, resulting in exponential growth \cite{WhiteGates}. Figure \ref{fig:radiativeIsland} shows good separation of disruptive and non-disruptive IRLMs in the $l_i$ and $q_{95}$ space used in reference \cite{disruptionsJET} to identify density limit disruptions. Although the plasmas discussed herein are expected to be far from the density limit \cite{Greenwald} as a result of the degraded confinement, they are likely near the radiative tearing instability limit which is theorized as the fundamental mechanism causing the density limit \cite{Gates2012}. This radiative tearing instability limit is expected qualitatively to scale with $l_i/q_{95}$.

$l_i/q_{95}$ and $d_{edge}$ are both good IRLM disruption predictors and are both correlated with $\rho_{q2}$ (the former by equation \ref{eq:q95RhoRelation}, and the latter by definition). Although this does imply some common underlying physics, $l_i/q_{95}$ and $d_{edge}$ differ from $\rho_{q2}$ by capturing information on the $q$-profile shape and the island width respectively. Both discriminate disruptive from non-disruptive IRLMs better than $\rho_{q2}$ alone, and do so by leveraging different physics. We conjecture that $l_i/q_{95}$ is a proxy for $\Delta'(w)$, as discussed earlier in this section and supported by earlier publications. The success of the $d_{edge}$ parameter might suggest a position-dependent critical 2/1 island width, where criticality depends on proximity to the unperturbed last closed flux surface. Alternatively, if the $n=1$ field is the result of multiple islands and not just the 2/1 island, as implicitly assumed in the definition of $d_{edge}$, then island overlap might result from $n=1$ islands of $m\geq2$ becoming closer and larger as $d_{edge}$ decreases. Island overlap is known to cause stochastic fields \cite{RebutIAEA}, and their effect on confinement might be sufficient to induce a disruption. Commonalities and differences between $l_i/q_{95}$ and $d_{edge}$ will be the subject of future work.

Other works have found parameters similar to $d_{edge}$ to be relevant to the
onset of the thermal quench, and are briefly listed here. In reference \cite{Izzo},
a stable Alcator C-Mod equilibrium is used as an initial condition for a numerical simulation of massive gas injection. It is found that by
uniformly distributing a high-Z gas in the edge, a 2/1 island is driven unstable and upon intersecting with the high-Z gas (i.e. when $d_{edge}$ becomes small), the 2/1 grows rapidly
followed by a 1/1 tearing mode that leads to a complete thermal quench. 
Similarly, a limited plasma is simulated in reference \cite{Sykes} with a current profile unstable to the 2/1 island, and the plasma current is ramped up to drive the 2/1 island towards the plasma edge. 
It is found that when the 2/1 island comes into contact with the limiter or a cold edge region (i.e. in both cases, when $d_{edge}$ is small), a rapid growth of the 
2/1 ensues followed by a 1/1 kink displacement of the core, ending in a disruption. 
Similarly, experimental results of disruptions induced by
EF penetration modes on COMPASS-C \cite{henderCompass} are attributed
to exceeding a threshold of $w/(a - r_s) > 0.7$, where $w$ and $r_s$ are the 2/1 island width and minor radius. This observation is thought to be due to the
2/1 island interacting with a 3/1 island.

Large 3-D fields at the plasma separatrix cause radial deformations of the edge plasma, leading to edge flux-surfaces 
(like the $q=2$ when $d_{edge}$ is small) intersecting the vessel \cite{homoclinicTangles}.  Further, some works suggest the existence of a stochastic layer within the
LCFS \cite{Evans,IzzoStoch}, which might facilitate
stochastization of the $2/1$ island when $d_{edge}$ is
sufficiently small. 

Separately, the radiation drive of tearing modes is
sensitive to edge proximity \cite{WhiteGates}, causing island growth
which reduces $d_{edge}$, and could then lead to the thermal quench
through one of the above mechanisms. 
Again, recall that our
interpretation of $d_{edge}$ at 20 ms prior to the disruption requires
the exponential growth observed in the $n=1$ $B_R$ signal to be
due to the 2/1 island growth, which is plausible but not confirmed due to lack of poloidal harmonic analysis in the
locked phase. Regardless of the theoretical interpretation,
$d_{edge}$ is a useful IRLM disruption predictor within 20 ms of the disruption for DIII-D.

\section*{Summary and conclusions}

Approximately 22,500 DIII-D plasma discharges were automatically
analyzed for the existence of initially rotating 2/1 locked
modes (IRLMs). The results of this analysis permits statistical
analysis of timescales, mode amplitude dynamics, effects of plasma
$\beta$ and major radius $R$, and disruptivity as a function of plasma
and mode properties. 

Timescales investigated suggest that a rotating 2/1 NTM that
will eventually lock rotates for $\sim200$ ms, decelerates from $f=2$
kHz to locked in $\sim15$ ms, and survives as a locked mode for
$\sim300$ ms (these values all represent the most frequent value in
their respective histograms). These timescales provide insight into
how to respond to a mode locking event, whether with a disruption avoidance approach (such as fast controlled shut-down, mode spin-up, or mode stabilization) or a disruption mitigation system (such as massive gas injection), depending on the response times of different systems and approaches.

Prior to disruption, the median $n=1$ perturbed field grows
consistent with exponential with an $e$-folding time between $\tau_g =[80,250]$ ms,
which might be due to exponential growth of the 2/1 IRLM. 

The parameter $l_i/q_{95}$ is shown to have a strong ability to
discriminate between disruptive and non-disruptive IRLMs, up to
hundreds of milliseconds before the disruption. 
As an example, the criterion $l_i/q_{95}>0.28$ in the presence of a detected IRLM
misses only 7\% of disruptions and produces 11\% false alarms, 
with at least 100 ms of warning time. 
$l_i/q_{95}$ might be related to the
free energy available to drive tearing growth.

$d_{edge}$ performs comparably to $l_i/q_{95}$ in its ability to discriminate disruptive IRLMs. A threshold below which IRLMs are
considered disruptive of $d_{edge}=9$ cm produces 
 4\% missed disruptions and 12\% false alarms,
with at least 20 ms of warning time. $d_{edge}$
is also observed to exhibit the best correlation with the IRLM survival
time. $d_{edge}$ might be a fundamental trigger of
the thermal quench, supported by similar observations by other
authors \cite{henderCompass,Sykes,Izzo,homoclinicTangles}. 

Future work will attempt to validate thermal quench onset mechanisms
in plasmas with locked modes, with the goal of a fundamental
understanding of locked mode disruptions, and thus how to avoid them.

\section{Acknowledgements}
This work was conducted in part under the DOE Grant DE-SC0008520. In addition, this work is supported by the U.S. Department of Energy, Office of Science, Office of Fusion Energy Sciences, using the DIII-D National Fusion Facility, a DOE Office of Science user facility under awards, DE-FG02-04ER54761$^1$, DE-FC02-04ER54698$^2$, and DE-FG02-92ER54139$^3$. DIII-D data shown in this paper can be obtained in digital format by following the links at https:\/\/fusion.gat.com\/global.D3D\_DMP.

The authors would like to thank E.J.~Strait and R.J.~Buttery for carefully reading the manuscript, and A.~Cole, J.~Hanson, E.~Kolemen, M.~Lanctot, L.~Lao, C.~Paz-Soldan, D.~Shiraki, and R.~Wilcox for fruitful conversations that contributed to this work. The authors would also like to thank S.~Flanagan and D.~Miller for their computer  expertise in developing this database, and M.~Brookman for assistance in preparing the electron cyclotron emission data.

\section*{References}
\label{sec:endnotes}

\bibliography{LMDatabase_Master_NF_v3}{}
\bibliographystyle{unsrt}

\section*{Appendix A - Mapping from radial field to perturbed island current}
\label{sec:appendixA}

A model was produced to map from the value of the $n=1$ major-radial
magnetic field $B_R$ at the external saddle loops to the perturbed
current carried by a circular cross-section toroidal current sheet. This model was inspired by a similar technique that accurately reproduced experimental magnetic signals in DIII-D \cite{shiraki}. The current sheet has major radius $R$ and minor radius $r_{q2}$, derived from equilibrium reconstructions 
using magnetics and Motional Stark Effect data. A shaping study was performed to test the impact of ellipticity on the $n=1$ external saddle
loop measurement. The impact was found to be relatively
small, introducing  corrections of less than 5\%.  Hence, it was
concluded that circular cross-sections are sufficient for modeling
external saddle loop signals for the sake of the present study.

\begin{figure}[t]
	\centering 
	\includegraphics[scale=0.04]{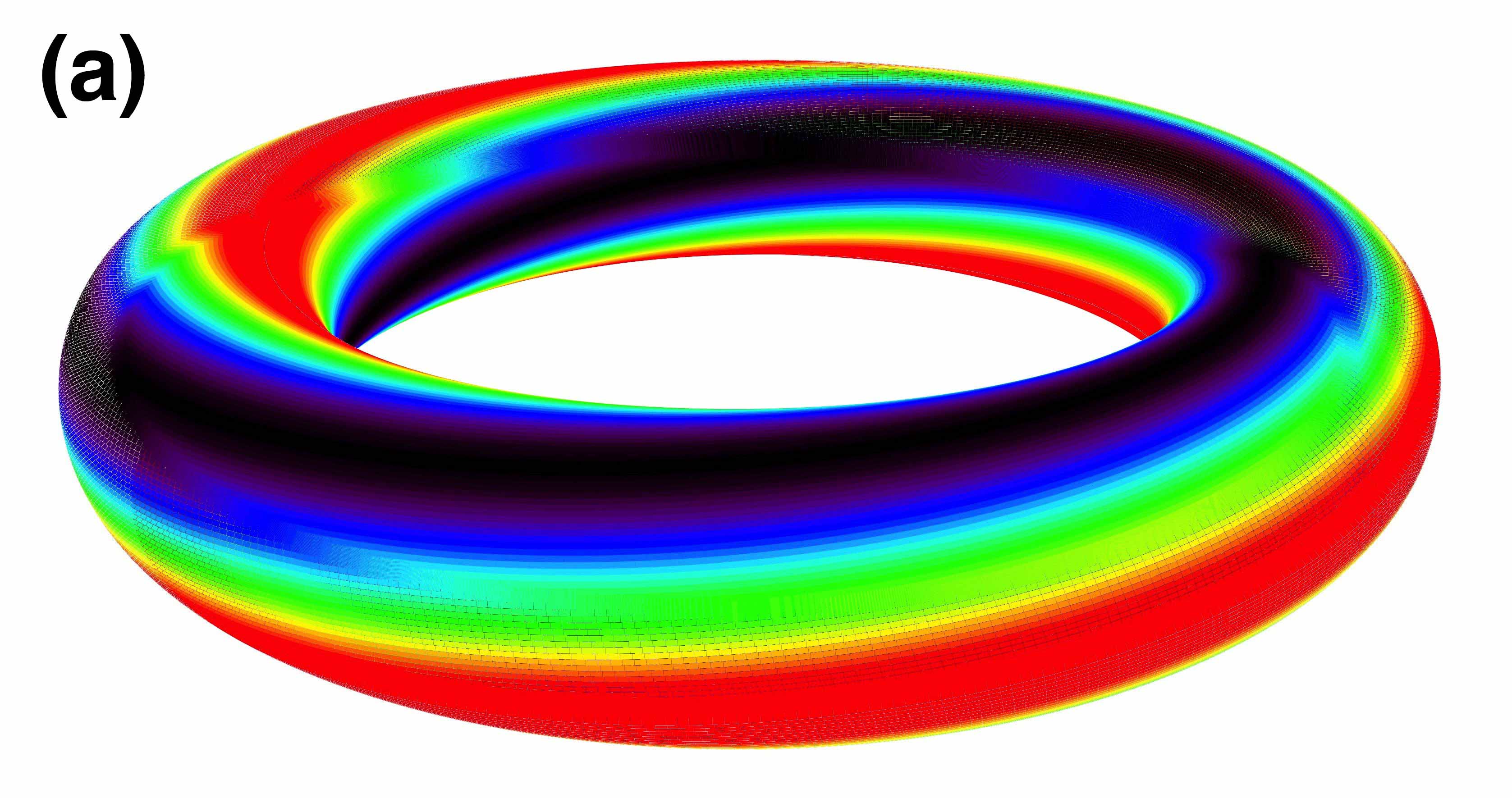} 
	\includegraphics[scale=0.04]{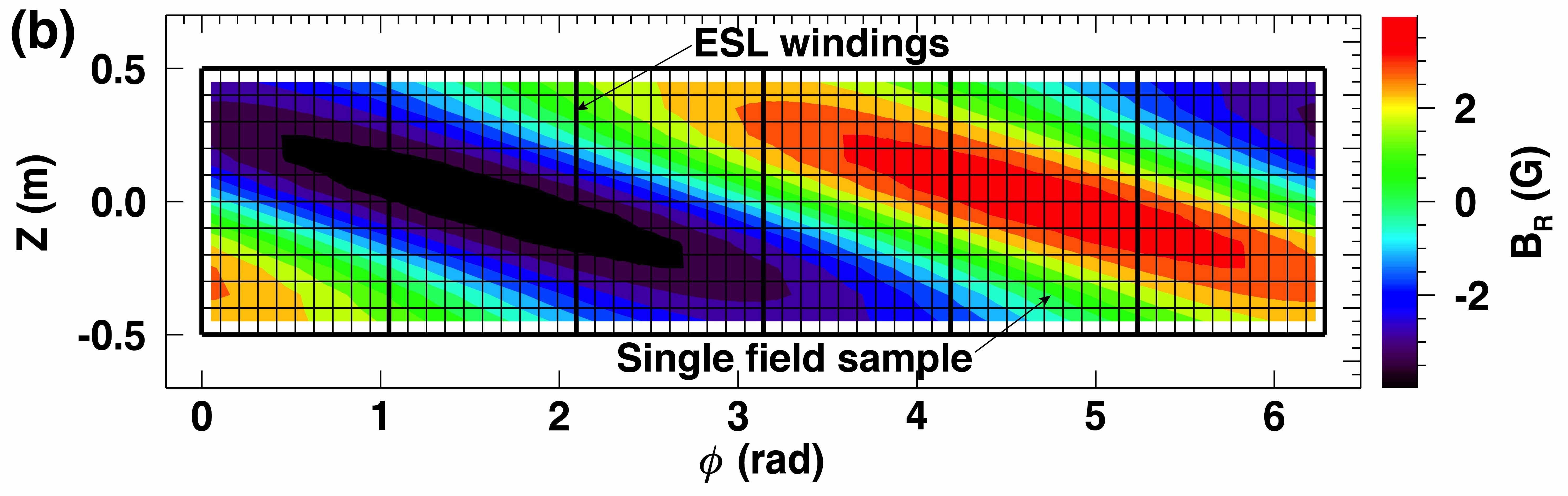}	
	\caption{
	(a)
	A 3D filament model of a 2/1 tearing mode used to map from the
        radial field at the external saddle loops $B_R$ to a perturbed
        island current $\delta I$.  This model, and figure, were inspired by \cite{shiraki}.
        (b)The radial field at
        the external saddle loops is shown in the color contour for a
        mode carrying 3.14 kA of $n=1$ current.  The thick black
        lines outline the six external saddle loops. The thin black
        lines define cells of equal  area, which are used to sample the field to compute an
        average over the loop.
        } \label{fig:tearing_Mode_3D}
\end{figure}

The modeled current sheet is discretized into $N$ helical wire
filaments with 2/1 pitch giving a toroidal
spacing of $360/N$ degrees. The currents are distributed among the wires to produce a
2/1 current perturbation as shown by the colors in
figure \ref{fig:tearing_Mode_3D}a (black represents no perturbed
current and red represents maximum perturbed current).

The field is calculated by numerically integrating the Biot-Savart Law
along each wire.  Figure \ref{fig:tearing_Mode_3D}b  shows the
resulting major-radial field $B_R$ as "seen" by the external saddle
loops, due to a 2/1 current distribution.  The color contour
shows a clear $n=1$ field distribution. Although the external saddle
loops have complete toroidal coverage, they have only
$\sim25\%$ poloidal coverage for the non-elongated, cylindrically approximated plasma. 

A single measurement for each saddle loop is estimated by
averaging the field at 100 sample points. Care was taken to ensure the area of each
sample point is identical, which simplifies the flux calculation. The
field samples are averaged, which is identical to calculating the total flux and dividing by saddle loop area $A_{sl}$:

\begin{equation}
\frac{ \sum_{i=1}^n B_{R,i}a_i } {A_{sl}} = \frac{a}{na} \sum_{i=1}^n B_{R,i} =\frac{1}{n} \sum_{i=1}^n B_{R,i}
\end{equation}

\noindent where $A_{sl}$ is the total area of one saddle loop, $a_i$ is the constant sample area
(i.e. $a_i=a$), $n$ is the number of sample regions per saddle loop,
and $B_{R,i}$ is the major-radial field at the $i^{th}$ sample
region.

The above averaging is done for each of the six saddle loops. The simulated signals are then pair-differenced, 
and a least squares fit is used to extract the amplitude of the $n=1$ 
component, in the same way that experimental data are analyzed 
in section \ref{sec:detectLocked}. 

\begin{figure}[t]
	\centering \includegraphics[scale=0.35]{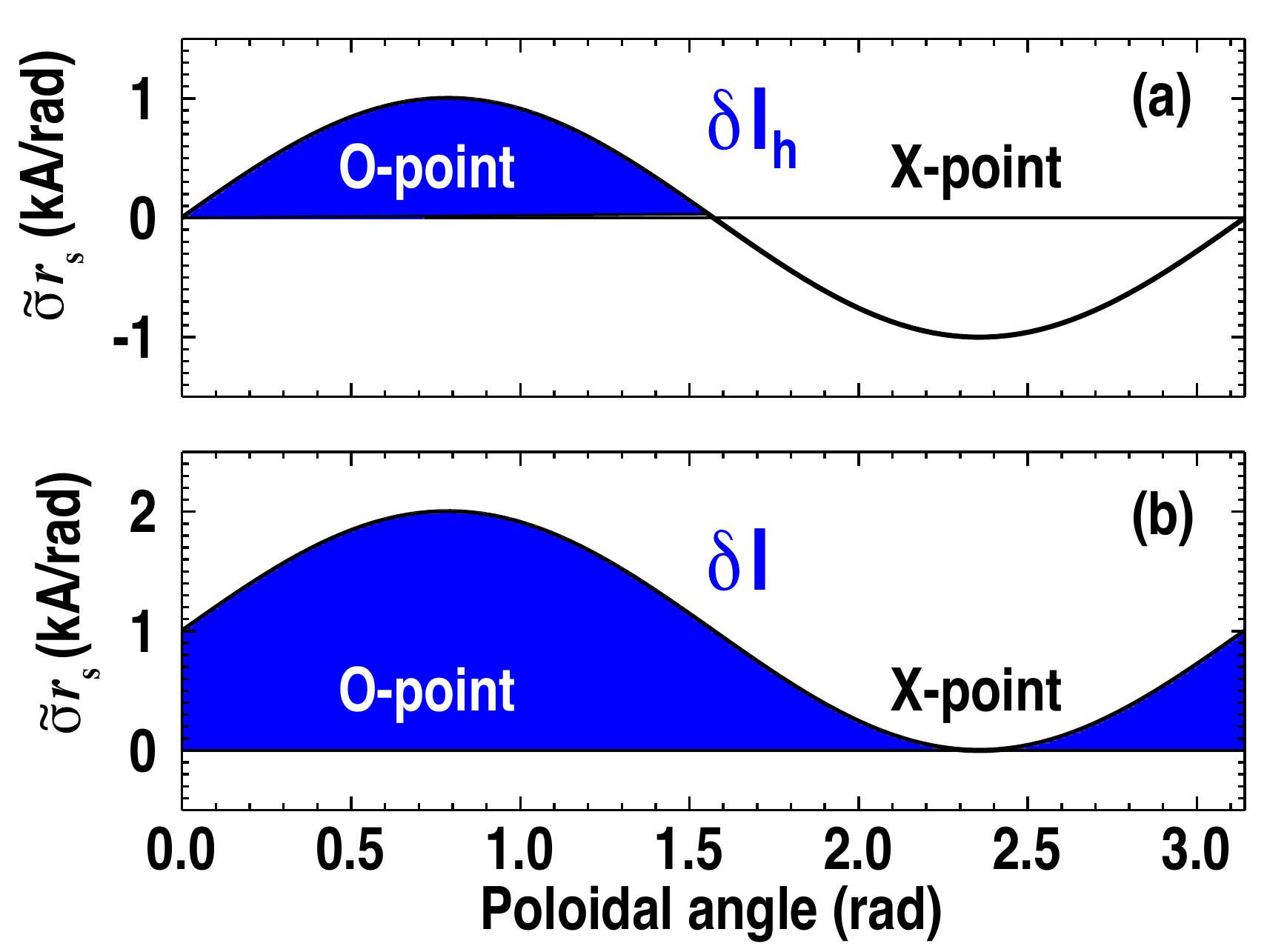} 
\caption{
(a) Half the perturbed island current $\delta I_h$ is shown by the blue
shaded region.  Here, it is assumed that the current perturbation is
sinusoidal about zero.  
(b) The total perturbed island current $\delta
I$ is shown by the blue shaded region.  Here, it is assumed that no
perturbed current flows at the X-point.  } \label{fig:deltaIDef}
\end{figure}

The island current, $\delta I$, is defined as the total current
deficit,  as seen in figure \ref{fig:deltaIDef}b. A
second definition of island current is useful for mapping to a cylindrical island width, being half the sinusoidal
perturbed current $\delta I_h$ (see figure \ref{fig:deltaIDef}a). $\delta I_h$ and $\delta I$ are  related by the equation
$\delta I = \pi \delta I_h$. The current distribution in
figure \ref{fig:deltaIDef}b can be seen as the
distribution in \ref{fig:deltaIDef}a, plus an axisymmetric current.
The non-axisymmetric fields produced by \ref{fig:deltaIDef}a and \ref{fig:deltaIDef}b are identical. 

\begin{figure}[t]
        \includegraphics[scale=0.3]{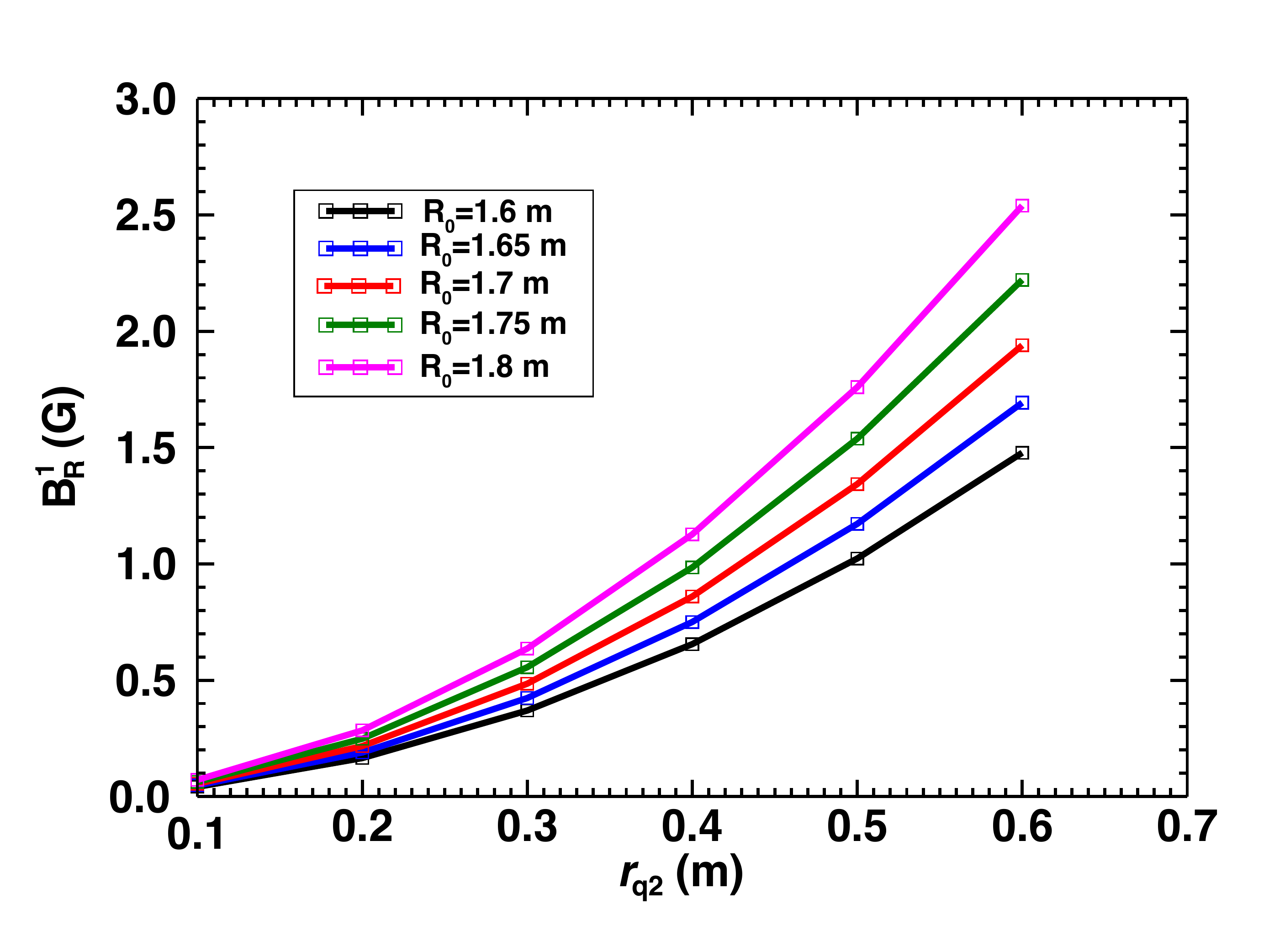} \caption{
        The $n=1$ major-radial field $B_R^1$ measured at the saddle loops as a function of the minor radius of the $q=2$ surface $r_{q2}$. 
        } \label{fig:BR_majorMinorR_Dependences}
\end{figure}

The $n=1$ $B_R$ component in the external saddle loops, $B_R^1$, was
then studied as a function of $R$ and $r_{q2}$ for fixed island
current $\delta I_h$ as seen in
figure \ref{fig:BR_majorMinorR_Dependences}. $B_R^1$ is found to be
quadratic in $r_{q2}$ in agreement with the cylindrical
approximation. As $R$ is increased, the leading order coefficient of
the quadratic is seen to increase as well. Defining the leading order coefficient
$\alpha(R)$ (which note is a function of $R$), we find the following relationship:

\begin{equation}
B_R^1 = \alpha(R) \delta I_h [r_{q2}]^2
\label{eq:BR1}
\end{equation}

\noindent where $B_R^1$ is in Gauss, $\delta I_h$ is in kA, $r_{q2}$ and
$R$ are in meters, and $\alpha$ has units of G/kA/m$^2$.  Evaluating
the change in $\alpha$ as a function of $R$, the dependence is found
to be well approximated by a second order polynomial $\alpha(R) = aR^2
+ bR + c$ where $a=19.15$ G/kA/m$^4$, $b=-50.40$ G/kA/m$^3$, and
$c=35.71$ G/kA/m$^2$.

With equation \ref{eq:BR1}, experimental $n=1$ measurements from the
external saddle loops are mapped to $\delta I_h$ by a simple
inversion of the equation and with major and minor radii provided by
EFIT informed by magnetics and Motional Stark Effect data:

\begin{equation}
\delta I_h = \frac{B_R^1}{\alpha(R) [r_{q2}]^2 }
\label{eq:delI}
\end{equation}

\section*{Appendix B - Mapping from perturbed island current to width}
\label{sec:appB}

Although the perturbed island current $\delta I$ is an intrinsic
island quantity that accounts for toroidicity, an island width was desirable for some studies. 
We will use $\delta I$ to map to an island width.

We solve for the radial field
at the $q=2$ surface $\tilde{B}_r$ by assuming cylindrical geometry, using the ordering $k_{\theta} = m/r \gg k_z = n/R$, and 
assuming vacuum solutions for the radial tearing eigenfunctions.  The
non-axisymmetric field $\tilde{\bv{B}}$ produced by the island is
expressed as,

\begin{equation}
\tilde{\bv{B}} = \curl \Psi \hat{\bv{z}}
\label{eq:Psi}
\end{equation}

\noindent where $\Psi$ is the perturbed flux function which is of the form $\Psi
= \psi(r) e^{im\theta}$ (note that the toroidal wavelength is assumed infinite here). We assume the perturbed current in the island
$\tilde{\bv{j}}$ to be a sheet current located at the $q=2$ surface
expressed as, 

\begin{equation}
\tilde{\bv{j}} = \tilde{\sigma} \sin(2 \theta)\delta (r_{q2} - r) \unit{z}
\label{eq:currentDensity}
\end{equation}

\noindent where $\tilde{\sigma}$ is constant with units (A/m), $\delta(...)$
represents the Dirac delta function, and the poloidal harmonic $m=2$
is implied.  Invoking Ampere's law where the time derivative of
$\bv{E}$ is neglected (the characteristic velocity of the system
$v \ll c$), and the equilibrium plasma current is neglected (i.e. searching for vacuum
solutions), we find

\begin{equation}
-\nabla^2 \Psi =  \mu_0 \tilde{\sigma} \sin(2\theta) \delta (r_{q2} - r) 
\end{equation}

Solving the  cylindrical Laplacian both inside and outside of the $q=2$
surface, requiring the solutions be continuous across $r_{q2}$, and
defining the  jump in the radial derivative with the radial integral
of $\tilde{\bv{j}}$,  we find

\begin{equation}
\Psi =  
\frac{\mu_0 \tilde{\sigma} }{4} \sin(2\theta)
\left \{
\begin{array}{lr} 
\frac{r^2}{r_{q2}}  &  \mathrm{for\:\:}  r <  r_{q2} \\
 & \\
\frac{r_{q2}^3 }{r^2}  &  \mathrm{for\:\:}  r \geq r_{q2}\\
\end{array}
\right .
\end{equation}

Using equation \ref{eq:Psi}, the field at the $q=2$ surface is then given by

\begin{equation}
\bv{\tilde{B}}_{q2} = \frac{\mu_0 \tilde{\sigma}}{2} \left [ \cos(2\theta) \unit{r} + \sin(2\theta) \unit{\theta} \right ]
\label{eq:modeRadialSigma}
\end{equation}

We now want to replace $\tilde{\sigma}$ with an expression for $\delta I$. We do so as follows (see
figure \ref{fig:deltaIDef}a),

\begin{equation}
\delta I_h \equiv \int_0^{\pi/2} \int_{r_{q2-}}^{r_{q2+}} \bv{\tilde{j}}\cdot d\bv{A}
\end{equation}

\noindent where $d\bv{A} =  r \: dr \: d\theta \: \unit{z}$, and $r_{q2\pm} = r_{q2} \pm w/2$. 
After substituting equation \ref{eq:currentDensity} and evaluating the 
integral, we find $\tilde{\sigma} = \delta I_h / r_{q2}$.

Finally, substituting this into equation \ref{eq:modeRadialSigma}, we have

\begin{equation}
\bv{\tilde{B}}_{q2}(\delta I) = \frac{\mu_0 \delta I_h}{2 r_{q2}} \left [ \cos(2\theta) \unit{r} + \sin(2\theta) \unit{\theta} \right ]
\label{eq:modeRadial}
\end{equation}

The cylindrical island width is given by \cite{Tokamaks},

\begin{equation}
w = c \left [ \frac{ 16 R \tilde{B}_r q^2 }{ m B_T dq/dr} \right
]^{1/2}
\label{eq:cylWidth}
\end{equation}

\noindent where $c$ is a toroidal correction for island widths measured at the outboard midplane,
$\tilde{B}_r$ is
the perturbed radial field at the $q=2$ surface, $q=2$ is the
local safety factor, $m=2$ is the poloidal harmonic,  $B_T$ is the
toroidal field at the magnetic axis, and $dq/dr$ is the
radial derivative of the safety factor evaluated at $r_{q2}$ on the
outboard midplane (the cylindrical expression for the safety factor $q=r B_T/ (R B_{\theta})$ was used here).  
Substituting $q=2$ and $m=2$,
and using the maximum of equation \ref{eq:modeRadial} for $\tilde{B}_r$, we have

\begin{equation}
w = c \left [ \frac{ 16 R }{ B_T dq/dr} \frac{ \mu_0 \delta I_h}{r_{q2}}  \right ]^{1/2}
\label{eq:hybridWidth}
\end{equation}

This expression for the island width was validated using the electron cyclotron emission
diagnostic \cite{ece} to observe the size of flattened regions in the temperature profile. Using nine islands, the toroidal correction factor is calibrated to
$c\approx8/15$. Using equation \ref{eq:hybridWidth}, we expect a $\sim25\%$ statistical error on island width estimates. This
expression for the island width is used for all plots of $w$ and
$d_{edge}$.

\end{document}